\documentclass[11pt, a4paper]{article}
\pdfoutput=1
\usepackage{jheppub}

\usepackage{amsmath}
\usepackage{mathtools}
\usepackage{epstopdf}

\usepackage[english,greek]{babel}
\usepackage{verbatim}

\usepackage{color}
\usepackage{pstricks,framed}

\usepackage{tensor}
\usepackage{amssymb}
\usepackage{amsthm}
\usepackage{amsfonts}
\usepackage{simplewick}

\usepackage{chngcntr}
\usepackage[retainorgcmds]{IEEEtrantools}

\usepackage[]{mciteplus}
\mciteSetBstSublistMode{n}
\mciteSetMidEndSepPunct{\quad$\bullet$\quad}{.}{\relax}
\usepackage{graphicx}

\counterwithin{equation}{section}
\title{{\color{black!50!blue}Large-Spin Expansions of GKP Strings}}
\author[a,b]{Emmanuel Floratos,}
\author[b]{George Georgiou}
\author[a,b]{and Georgios Linardopoulos}
\affiliation[a]{Department of Physics, National and Kapodistrian University of Athens,\\
Zografou Campus, 157 84, Athens, Greece. \\}
\affiliation[b]{Institute of Nuclear and Particle Physics, N.C.S.R. "Demokritos",\\
153 10, Agia Paraskevi, Greece. \\}
\emailAdd{mflorato@phys.uoa.gr}
\emailAdd{georgiou@inp.demokritos.gr}
\emailAdd{glinard@inp.demokritos.gr}
\abstract{We demonstrate that the large-spin expansion of the energy of Gubser-Klebanov -Polyakov (GKP) strings that rotate in $\mathbb{R}\times\text{S}^2$ and AdS$_3$ can be expressed in terms of Lambert's W-function. We compute the leading, subleading and next-to-subleading series of exponential corrections to the infinite-volume dispersion relation of GKP strings that rotate in $\mathbb{R}\times\text{S}^2$. These strings are dual to the long $\mathcal{N} = 4$ SYM operators $\text{Tr}\left[\Phi\,\mathcal{Z}^m \, \Phi \, \mathcal{Z}^{J-m}\right] + \ldots$ and provide their scaling dimensions at strong coupling. We also show that the strings obey a short-long (strings) duality. For the folded GKP strings that spin inside AdS$_3$ and are dual to twist-2 operators, we confirm the known formulas for the leading and next-to-leading coefficients of their anomalous dimensions and derive the corresponding expressions for the next-to-next-to-leading coefficients.}
\arxivnumber{1311.5800}
\begin{document}
\selectlanguage{english}
\maketitle\flushbottom\normalsize
%
%\newpage
%\tableofcontents
%
\section[Introduction]{Introduction}
\renewcommand{\thefootnote}{\arabic{footnote}}
According to the standard lore of AdS/CFT correspondence \cite{Maldacena97, GubserKlebanovPolyakov98, Witten98a}, the partition functions of type IIB string theory on AdS$_5 \times \text{S}^5$ and $\mathcal{N} = 4$, $SU(N)$ super Yang-Mills (SYM) theory are equal. The field-operator correspondence of AdS/CFT states that every operator of $\mathcal{N} = 4$ SYM has a dual string state in IIB string theory on AdS$_5 \times \text{S}^5$ and every boundary observable (e.g. operator scaling dimensions, correlation functions, scattering amplitudes, etc.) possesses a dual observable in the bulk. Therefore, in order to establish the validity of AdS/CFT \cite{MAGOO99, DHokerFreedman02}, the spectra of scaling dimensions of the two theories should match (among other things and at least in the planar limit \cite{KristjansenStaudacherTseytlin09, Beisertetal12}). \\
\indent The difficulty with the program of matching the spectra on both sides of the duality has traditionally been linked to the insufficient knowledge of type IIB String Theory on $\text{AdS}_5 \times \text{S}^5$ backgrounds. A breakthrough in this direction was made in 2002 by Gubser, Klebanov and Polyakov (GKP) \cite{GubserKlebanovPolyakov02}, who proposed to study closed semiclassical strings, spinning, rotating or pulsating in $\text{AdS}_5 \times \text{S}^5$, in order to obtain the (anomalous) scaling dimensions of their dual SYM operators at strong coupling, a regime where all perturbative calculations from the gauge theory side typically break down \cite{Tseytlin11}. \\
\indent The paper of Gubser, Klebanov and Polyakov \cite{GubserKlebanovPolyakov02} contains three prototype string settings for which the energy-spin relation is calculated: (I) a folded closed string rigidly rotating at the equator of $\text{S}^3$ of $\text{AdS}_5$, (II) a folded closed string rigidly rotating around the pole of $\text{S}^2 \subset \text{S}^5$ and (III) a closed string pulsating inside $\text{AdS}_3$. Each of these string configurations is dual to a gauge-invariant operator of $\mathcal{N} = 4$ SYM, the (anomalous) scaling dimensions of which at strong coupling are given by the energy of the corresponding closed string state. \\
\indent Case (I) has been thoroughly analyzed over the past years. The key observation of GKP was that the energy minus the spin angular momentum of a long folded closed string that rotates inside AdS$_3$, scales as the logarithm of its (large) spin, behavior already familiar from the study of anomalous dimensions of twist-2 Wilson operators in perturbative QCD. Being able to reproduce this calculation within $\mathcal{N} = 4$ SYM, GKP asserted that their string is dual to the twist-2 operators of $\mathcal{N} =4$ SYM, Tr$\left[\mathcal{Z} \, \mathcal{D}_+^S \, \mathcal{Z}\right] + \ldots$ \footnote{We denote $\mathcal{Z}$, $\mathcal{W}$, $\mathcal{Y}$ the three complex scalars of $\mathcal{N} =4$ SYM composed out of its six real scalars $\phi$. Also $\mathcal{D}_+ = \mathcal{D}_0 + \mathcal{D}_3$, $\mathcal{D}_- = \mathcal{D}_1 + \mathcal{D}_2$ stand for the light-cone derivatives. The dots in a trace operator generally denote terms that are built by permuting trace fields $\mathcal{Z}$ and impurities, $\mathcal{W}$, $\mathcal{Y}$, $\mathcal{D}_\pm$.} \\
\indent Twist operators of large spin $S$ originally came up in a QCD context where their anomalous scaling dimensions were shown to be responsible for the (logarithmic) violation of Bjorken scaling in deep inelastic scattering (DIS). For twist-2 they have been calculated\footnote{For more references and a brief historical perspective, see \cite{AxenidesFloratosKehagias03}.} at one-loop \cite{GeorgiPolitzer74, GrossWilczek74}, two-loops \cite{FloratosRossSachrajda77, FloratosRossSachrajda78a, CurciFurmanskiPetronzio80, FloratosKounnasLacaze81} and three loops \cite{MochVermaserenVogt04a, MochVermaserenVogt04b}. As it turns out, the QCD results may be used to extract the corresponding anomalous dimensions in perturbative $\mathcal{N} = 1,2,4$ SYM theory (see e.g. \cite{AxenidesFloratosKehagias03}). For $\mathcal{N} = 4$ SYM the anomalous dimensions of twist-2 operators Tr$\left[\mathcal{Z} \, \mathcal{D}_+^S \, \mathcal{Z}\right] + \ldots$ have thus been calculated to one-loop \cite{KotikovLipatov01}, two-loops \cite{KotikovLipatovVelizhanin03} and, using transcedentality, to three-loops \cite{KotikovLipatovOnishchenkoVelizhanin04} for large-spin $S$ and weak 't Hooft coupling $\lambda$. To wit, the following logarithmic behavior is obtained, also known as Sudakov scaling:
\begin{IEEEeqnarray}{c}
\gamma\left(S\,,g\right) = \Delta - \left(S + 2\right) = f(g) \ln S + \ldots \; , \qquad g = \frac{\sqrt{\lambda}}{4\pi}\,, \quad \lambda = g_{\text{\tiny{YM}}}^2\,N. \label{AnomalousDimensions1}
\end{IEEEeqnarray}
$f(g)$ is known as the cusp anomalous dimension or the universal scaling function of $\mathcal{N} = 4$ SYM and has a history of its own with analytic expressions obtained by solving the Beisert-Eden-Staudacher (BES) equation at weak \cite{EdenStaudacher06, BeisertEdenStaudacher06} and strong coupling \cite{BassoKorchemskyKotanski07, KostovSerbanVolin08}. The latter is found to agree at two loops with the cusp anomalous dimension that is calculated from the folded closed AdS$_3$ string when quantum corrections are included \cite{FrolovTseytlin02, RoibanTirziuTseytlin07, RoibanTseytlin07}. \\
\indent Alternative techniques may provide the anomalous dimensions of twist-2 operators of $\mathcal{N} = 4$ SYM at weak coupling up to five loops. By solving the Baxter equation analytically, three-loop \cite{KotikovRejZieme08} and four-loop \cite{BeccariaBelitskyKotikovZieme09} expressions are obtained. By including wrapping corrections after three-loops, the anomalous dimensions to four and five-loops have been computed in \cite{BajnokJanikLukowski08, LukowskiRejVelizhanin09}. \\
\indent At strong coupling, the structure\footnote{For recent developments on the structure of 3-point correlators involving twist-$J$ operators both at weak and strong coupling, see \cite{Georgiou10, Georgiou11} and references therein.} of the large-spin expansion of anomalous scaling dimensions of Tr$\left[\mathcal{Z} \, \mathcal{D}_+^S \, \mathcal{Z}\right] + \ldots$ is identical to the perturbative one \eqref{AnomalousDimensions1}, albeit with different coefficients $f(\sqrt{\lambda})$:
\small\begin{IEEEeqnarray}{c}
\gamma\left(S\,,\lambda\right) = f \ln(S/\sqrt{\lambda}) + \sum_{n = 1}^\infty f_{nn} \frac{\ln^n(S/\sqrt{\lambda})}{S^n} + \sum_{n = 1}^\infty f_{nn-1} \frac{\ln^n(S/\sqrt{\lambda})}{S^{n+1}} + \sum_{n = 0}^\infty \frac{f_n}{S^n} + \ldots \label{AnomalousDimensions2} \qquad
\end{IEEEeqnarray}\normalsize
It is a daunting task to calculate all the coefficients of \eqref{AnomalousDimensions2} at strong coupling. Theoretically, one may obtain them all by means of the thermodynamic Bethe ansatz (TBA) \cite{BombardelliFioravantiTateo09}. Alternatively, one may calculate quantum corrections to the AdS$_3$ GKP-string \cite{Beccariaetal09, Beccariaetal10a}. \\
\indent In a recent paper, the authors of \cite{GeorgiouSavvidy11} succeeded in calculating all the classical leading and subleading terms of series \eqref{AnomalousDimensions2} at strong coupling, by introducing an iterative method that can potentially supply all the classical terms at an arbitrary subleading order. The current work extends the method of \cite{GeorgiouSavvidy11} to examples other than the classic GKP case (I).\footnote{At this point let us mention that an interesting generalization of this calculation would be to calculate the anomalous dimensions of twist-2 operators in the context of a generalized Yang-Mills theory, such as \cite{Savvidy10b}.} In their original treatment \cite{GubserKlebanovPolyakov02}, Gubser, Klebanov and Polyakov gave the following formula for the anomalous dimensions of the $\mathcal{N} = 4$ SYM operator that is dual to the $\mathbb{R}\times\text{S}^2$ closed and folded string (II):
\begin{IEEEeqnarray}{c}
E - J = \frac{2\sqrt{\lambda}}{\pi}, \quad J,\,\lambda \rightarrow \infty. \label{AnomalousDimensions3}
\end{IEEEeqnarray}
For finite yet large $J$, this expression receives subleading exponential contributions. In this paper we are interested in calculating all of the subleading in $J$ terms of the series \eqref{AnomalousDimensions3}. \\
\indent GKP case (II) also turns out to have been extensively studied over the past years. Some initial considerations based on the GKP model \cite{GubserKlebanovPolyakov02} may be found in \cite{FrolovTseytlin02, ChongLuPope04}. Closed folded single-spin strings rotating on $\mathbb{R}\times\text{S}^2$ can be decomposed into more elementary string theory excitations, known as giant magnons. These are open, single-spin strings rotating in $\text{S}^2 \subset \text{S}^5$, identified in 2006 by Hofman and Maldacena \cite{HofmanMaldacena06} as the string theory duals of $\mathcal{N} = 4$ SYM magnon excitations. The energy-spin relation of one giant magnon of angular extent $\Delta\varphi$ is:
\begin{IEEEeqnarray}{c}
E - J = \frac{\sqrt{\lambda}}{\pi} \, \left|\sin\frac{\Delta\varphi}{2}\right|, \quad J,\,\lambda \rightarrow \infty, \label{GiantMagnon1}
\end{IEEEeqnarray}
where $\Delta\varphi = p$ is the dual magnon's momentum. Superimposing two such giant magnons of maximum angular extent $\Delta\varphi = \pi$ gives the GKP formula \eqref{AnomalousDimensions3}.\\
\indent As we have just mentioned, giant magnons are dual to magnons, the elementary spin chain excitations. The exact infinite-volume magnon dispersion relation, valid at weak and strong coupling, was found by Beisert in \cite{Beisert05b} by considering the centrally extended superalgebra $\mathfrak{su}\left(2|2\right)_c \oplus \mathfrak{su}\left(2|2\right)_c \subset \mathfrak{psu}\left(2,2|4\right)$ :
\begin{IEEEeqnarray}{c}
E - J = \sqrt{1 + \frac{\lambda}{\pi^2}\sin^2\left(\frac{p}{2}\right)}, \quad J\rightarrow\infty. \label{GiantMagnon2}
\end{IEEEeqnarray}
This result reproduces both the string \eqref{GiantMagnon1} and the perturbative gauge-theory results. \\
\indent On the side of $\mathcal{N} = 4$ SYM, it has long been known that the dilatation operator has the form of an integrable spin chain Hamiltonian\footnote{The compact $\mathfrak{su}\left(2\right)$ sector of $\mathcal{N} = 4$ SYM consists of the single-trace operators $\text{Tr}\left[\mathcal{Z}^{L}\Phi^{L - M}\right]$, where $\Phi \in \left\{\mathcal{W}, \mathcal{Y}\right\}$. It is dual to strings rotating in $\mathbb{R}\times\text{S}^3 \subset \text{AdS}_5\times\text{S}^5$ and its one-loop dilatation operator is given by the Hamiltonian of the ferromagnetic XXX$_{1/2}$ Heisenberg spin chain. The other closed sector of relevance to our paper is the non-compact $\mathfrak{sl}\left(2\right)$ sector composed out of the twist-$J$ operators $\text{Tr} \left[\mathcal{D}_+^{S_1} \Phi \; \mathcal{D}_+^{S_2} \Phi \ldots \mathcal{D}_+^{S_J} \, \Phi\right]$, where $S_1 + S_2 + \ldots + S_J = S$ and $\Phi$ is any of the three complex scalar fields of $\mathcal{N} = 4$ SYM. This sector is dual to strings rotating in $\text{AdS}_3\times\text{S}^1$ and its dilatation operator is given by the Hamiltonian of the ferromagnetic XXX$_{-1/2}$ Heisenberg spin chain.} which can be diagonalized by means of the Bethe ansatz (BA) \cite{MinahanZarembo03}. However, and due to their being asymptotic in nature, the BA equations can only reproduce the correct form of anomalous dimensions only when the length $L$ of the spin chain is infinite or larger than the loop order $L$. At and above this \textit{critical} loop-order $L$, virtual particles start circulating around the spin chain (as the range of spin-chain interactions then exceeds its length) and wrapping corrections have to be taken into account. \\
\indent Indeed, the inefficiency of BA has been noted in both gauge \cite{KotikovLipatovRejStaudacherVelizhanin07} and string theories \cite{Schafer-NamekiZamaklarZarembo05, Schafer-NamekiZamaklar05}. The wrapping effects that appear at the critical loop-order have the form of exponentially small corrections to the anomalous dimensions, as noticed in \cite{Schafer-Nameki06, Schafer-NamekiZamaklarZarembo06}. An important theoretical issue is therefore the calculation of the exact anomalous dimensions of unprotected operators of $\mathcal{N} = 4$ SYM that have finite size $L$.\\
\indent To this end, it was proposed in \cite{AmbjornJanikKristjansen05} that the thermodynamic Bethe ansatz (TBA) \cite{Zamolodchikov89a}\,\footnote{For more, see the review \cite{Bajnok10}.} can correctly account for the wrapping interactions. The Y-system \cite{GromovKazakovVieira09a} also accounts for wrapping corrections. On the string theory side, one equivalently calculates exponential corrections to the corresponding giant magnon dispersion relation \eqref{GiantMagnon1}. The following result was first derived by Arutyunov, Frolov and Zamaklar in \cite{ArutyunovFrolovZamaklar06}:\footnote{Actually, this is the gauge-independent result of \cite{AstolfiForiniGrignaniSemenoff07}, which coincides with that of AFZ \cite{ArutyunovFrolovZamaklar06} in the temporal gauge.}
\small\begin{IEEEeqnarray}{ll}
E - J = &\frac{\sqrt{\lambda}}{\pi} \, \left|\sin\frac{p}{2}\right| \, \Bigg\{1 - 4\,\sin^2\frac{p}{2}\,e^{- 2 - 2\pi J/\sqrt{\lambda} \left|\sin\frac{p}{2}\right|} - 8\,\sin^2\frac{p}{2}\,\bigg[6\,\cos^2\frac{p}{2} + \frac{1}{2} + \nonumber \\
& + \left(6\,\cos^2\frac{p}{2} - 1\right)\frac{2\pi J} {\sqrt{\lambda}\left|\sin\frac{p}{2}\right|} + \cos^2\frac{p}{2}\,\left(\frac{2\pi J}{\sqrt{\lambda}\left|\sin\frac{p}{2}\right|}\right)^2\bigg]e^{- 4 - 4\pi J/\sqrt{\lambda}\left|\sin\frac{p}{2}\right|} + \ldots\Bigg\}. \qquad \label{GiantMagnon3}
\end{IEEEeqnarray} \normalsize
Astolfi, Forini, Grignani and Semenoff have proven in \cite{AstolfiForiniGrignaniSemenoff07} that, when placed upon an orbifold, giant magnons are completely independent of any gauge parameter. The first two terms of the same result \eqref{GiantMagnon3} have been found by the algebraic curve method in \cite{MinahanSax08}. They have also been obtained by using L\"{u}scher's perturbative method \cite{Luscher85a, KlassenMelzer90a} at strong coupling \cite{JanikLukowski07, HellerJanikLukowski08, GromovSchafer-NamekiVieira08a}. \\
\indent By using the relation between strings on S$^2$ and the sine-Gordon model \cite{Pohlmeyer75}, Klose and McLoughlin \cite{KloseMcLoughlin08} have obtained the following leading terms of the series \eqref{GiantMagnon3}:
\small\begin{IEEEeqnarray}{ll}
E - J &= \frac{\sqrt{\lambda}}{\pi} \, \sin\frac{p}{2} \, \Bigg\{1 - 4\,\sin^2\frac{p}{2}\,e^{-L_{\text{eff}}}\bigg[1 + 2\,L_{\text{eff}}^2\,\cos^2\frac{p}{2}\,e^{-L_{\text{eff}}} + 8\,L_{\text{eff}}^4\,\cos^4\frac{p}{2}\,e^{-2L_{\text{eff}}} + \nonumber \\
&+ \frac{128}{3}\,L_{\text{eff}}^6\,\cos^6\frac{p}{2}\,e^{-3L_{\text{eff}}} + \frac{800}{3}\,L_{\text{eff}}^8\,\cos^8\frac{p}{2}\,e^{-4L_{\text{eff}}} + \frac{9216}{5}\,L_{\text{eff}}^{10}\,\cos^{10}\frac{p}{2}\,e^{-5L_{\text{eff}}} + \ldots\bigg]\Bigg\}, \qquad \label{GiantMagnon4}
\end{IEEEeqnarray} \normalsize
where $L_{\text{eff}} \equiv L / \sin p/2$ is the effective length and $L$ is the spatial periodicity of the spin chain. Giant magnons have been generalized to $\beta$-deformed backgrounds \cite{ChuGeorgiouKhoze06, BobevRashkov06}, TsT-transformed AdS$_5\times\text{S}^5$ \cite{BykovFrolov08, AhnBozhilov10} and AdS$_4/\text{CFT}_3$ \cite{GaiottoGiombiYin08, GrignaniHarmarkOrselli08}. \\
\indent In this paper we study the dispersion relations of the following (long) operators of $\mathcal{N} = 4$ SYM:
\begin{IEEEeqnarray}{c}
\mathcal{O}_S = Tr\left[\mathcal{Z} \, \mathcal{D}_+^S \, \mathcal{Z}\right] + \ldots \quad \& \quad \mathcal{O}_J = \text{Tr}\left[\Phi\mathcal{Z}^m \, \Phi \, \mathcal{Z}^{J-m}\right] + \ldots, \label{Operators1}
\end{IEEEeqnarray}
for large values of the 't Hooft coupling constant $\lambda$. These operators are dual to cases (I) and (II) of GKP strings that were discussed above. Our work is motivated by the interesting properties of twist-2 operators that originally appeared in the context of deep inelastic scattering in QCD and have been studied intensively ever since. Twist-2 operators are also present in $\mathcal{N} = 4$ super Yang-Mills theory and the AdS/CFT correspondence permits to study them in the strong coupling regime. Our investigation aims to shed light on the structure of the corresponding strong-coupling, large-spin, semiclassical dispersion relations, with a view to gaining insight into their weak-coupling, small-spin and quantum generalizations. A reorganization of these relations might lead to better-recognizable structures that could be useful, especially when one is trying to derive series \eqref{AnomalousDimensions2}-\eqref{GiantMagnon3} with other integrability methods such as L\"{u}scher corrections, the TBA or the Y-system.\\
\indent We propose a systematic method, based on the Lagrange-B\"{u}rmann inversion formula, for the order-by-order inversion of a certain class of functions that are related to elliptic integrals. Next, we apply this method in order to express the dispersion relations of operators \eqref{Operators1} as sums of Lambert's W-functions. The latter are analytic in their arguments and therefore have the property of collecting infinitely many terms that were previously unbeknown. We emphasize that these previously unknown terms have not yet been obtained by any other integrability method, such as L\"{u}scher corrections (where only the first two terms are known), the TBA or the Y-system. Within the context of AdS/CFT correspondence, our analysis could have repercussions in the dispersion relations of giant magnons, spiky strings, spinning membranes, or strings rotating in $\beta$-deformed and TsT-transformed backgrounds. Our technique can also be applied to the solution of the renormalization group equations up to any loop-order. For more, the reader is referred to the discussion section \ref{Discussion}. \\
\indent Our paper is organized as follows. We begin with a brief summary of our findings in section \ref{StrongSummary}. In section \ref{GKPStringII} we derive the single-spin string solution of GKP that rotates in $\mathbb{R}\times\text{S}^2$ by minimizing the energy functional of a generic $\mathbb{R}\times\text{S}^2$ string configuration and show that it obeys a duality of the short-long type. Section \ref{SpinFunctionII} contains our calculation of the exponential corrections to the anomalous dimensions of long $\mathbb{R}\times\text{S}^2$ strings. We start from a $2 \times 2$ system of equations
\begin{IEEEeqnarray}{c}
\mathcal{E} = d\left(x\right) \ln x + h\left(x\right) \label{Method1} \\
\mathcal{J} = c\left(x\right) \ln x + b\left(x\right), \label{Method2}
\end{IEEEeqnarray}
\noindent where $\mathcal{E}$ and $\mathcal{J}$ are the string's energy and spin, $x$ is a parameter depending on the angular velocity of the string and $d\left(x\right)$, $h\left(x\right)$, $c\left(x\right)$, $b\left(x\right)$ are some known power series of $x$. We invert equation \eqref{Method2} for the inverse spin function $x = x\left(\mathcal{J}\right)$, then plug it back into \eqref{Method1} in order to obtain the anomalous dimensions $\gamma \equiv \mathcal{E} - \mathcal{J} = \gamma\left(\mathcal{J}\right)$. The main thrust of our paper is showing that the dispersion relation $\gamma\left(\mathcal{J}\right)$ can be written in terms of Lambert's W-function:
\begin{IEEEeqnarray}{c}
W\left(z\right)\,e^{W\left(z\right)} = z \Leftrightarrow W\left(z\,e^z\right) = z. \label{LambertDefinition1}
\end{IEEEeqnarray}
We obtain formulas for all the leading, subleading and next-to-subleading coefficients for both long folded (angular velocity $\omega > 1$) and fast circular (angular velocity $\omega < 1$) strings in $\mathbb{R}\times\text{S}^2$. In sections \ref{GKPStringI}--\ref{SpinFunctionI}, our analysis is repeated for long folded strings spinning (with angular velocity $\omega > 1$) in $\text{AdS}_3$ where again we derive expressions for the leading, subleading and next-to-subleading coefficients of anomalous dimensions. A discussion of our results can be found in section \ref{Discussion}. Appendix \ref{LambertAppendix} is a brief introduction to Lambert's W-function. Appendix \ref{EllipticFunctionsAppendix} contains the definitions and some useful formulas of elliptic integrals and functions. In appendix \ref{Short-Slow-StringsAppendix} we briefly consider short/slow-spin strings and in appendix \ref{Long-Fast-StringsAppendix} we have collected our symbolic computations on the long and fast GKP strings of our paper. \\
\section[Summary of Results]{Summary of Results \label{StrongSummary}}
In this section we summarise our basic results.
\subsection[Long Strings in $\mathbb{R}\times\text{S}^2$]{Long Strings in $\mathbb{R}\times\text{S}^2$}
The $\mathbb{R}\times\text{S}^2$ string solutions that we will be considering are dual to the following gauge theory operators: $\text{Tr}\left[\Phi\mathcal{Z}^m \, \Phi \, \mathcal{Z}^{J-m}\right] + \ldots$, where the dots denote appropriate weighted permutations of the complex scalar fields $\mathcal{Z}$, $\Phi \in \left\{\mathcal{W}, \mathcal{Y}\right\}$ and $J \gg \sqrt{\lambda}$. By computing the energy of the GKP string, we obtain the strong coupling value of scaling dimensions of the aforementioned operators. The result can be expressed in terms of Lambert's function $W\left(\pm 8 \mathcal{J} e^{-2\mathcal{J} - 2}\right)$ as follows:
\small\begin{IEEEeqnarray}{l}
\mathcal{E} - \mathcal{J} = 1 - \frac{1}{4\mathcal{J}}\left(2 W + W^2\right) - \frac{1}{16\mathcal{J}^2}\left(W^2 + W^3\right) -
\frac{1}{256\mathcal{J}^3}\frac{W^3\left(11\,W^2 + 26\,W + 16\right)}{1 + W} + \ldots, \qquad \label{AnomalousDimensions0}
\end{IEEEeqnarray}
\begin{IEEEeqnarray}{l}
\bullet \ \text{leading terms: } - \frac{1}{4\mathcal{J}}\left(2 W + W^2\right) = \sum_{n = 1}^{\infty} \mathfrak{a}_n \, \mathcal{J}^{n - 1} \, \left(e^{-2\mathcal{J} - 2}\right)^{n}. \nonumber \\[6pt]
\bullet \ \text{subleading terms: } - \frac{1}{16\mathcal{J}^2}\left(W^2 + W^3\right) = \sum_{n = 2}^{\infty} \mathfrak{b}_n \, \mathcal{J}^{n - 2} \, \left(e^{-2\mathcal{J} - 2}\right)^{n}. \nonumber \\[6pt]
\bullet \ \text{next-to-subleading terms: } - \frac{1}{256\mathcal{J}^3}\frac{W^3\left(11\,W^2 + 26\,W + 16\right)}{1 + W} = \sum_{n = 3}^{\infty} \mathfrak{c}_n \, \mathcal{J}^{n - 3} \, \left(e^{-2\mathcal{J} - 2}\right)^{n}. \nonumber
\end{IEEEeqnarray} \normalsize
where $\mathcal{E} \equiv \pi\,E/2\sqrt{\lambda}$ and $\mathcal{J} \equiv \pi\,J/2\sqrt{\lambda}$. The plus sign in the argument of Lambert's W-function corresponds to closed and folded strings (angular velocity $\omega>1$) and the minus sign to circular strings (angular velocity $\omega<1$). Upon expansion of Lambert's W-function, the second, third and fourth term on the r.h.s.\ of \eqref{AnomalousDimensions0} provide three infinite series of coefficients which completely determine the leading, subleading and next-to-subleading contributions to the large-$J$ finite-size corrections to the dispersion relation of a closed folded single-spin string rotating on S$^2$. The precise expressions for these infinite series can be found in equations \eqref{AnomalousDimensions8}, \eqref{AnomalousDimensions12} and \eqref{AnomalousDimensions15}. One can argue that all finite-size corrections ($N^k$-subleading terms) can be written in terms of Lambert's W-function. Note that all (super-leading) terms of the form $\mathcal{J}^{n}(e^{-2 \mathcal{J}-2})^n$ are absent from expansion \eqref{AnomalousDimensions0}. Effectively, our calculation provides the finite-size corrections to the dispersion relation of a single-magnon state with maximal momentum $p = \pi$, the dual operator of which is:
\begin{IEEEeqnarray}{c}
\widetilde{\mathcal{O}}_J = \sum_m e^{im\pi} \left|\ldots\mathcal{Z} \mathcal{Z}\Phi(m)\mathcal{Z}\mathcal{Z}\ldots\right>\,, \quad J\rightarrow\infty.
\end{IEEEeqnarray}
These corrections are exactly equal to one-half the finite-size corrections \eqref{AnomalousDimensions0} to the energy of a closed folded single-spin string that rotates on S$^2$. \\
\indent We have also found a duality between short and long strings. For each solution of energy $E$ and spin $J$, there exists a dual solution whose energy $E'$ and spin $J'$ are related to the original by equations \eqref{Short-LongII1}--\eqref{Short-LongII2}. \\
\subsection[Long Strings in AdS$_3$]{Long Strings in AdS$_3$}
Secondly, we will consider the classical GKP solution rotating in an AdS$_3$ subspace of AdS$_5$. These strings are dual to twist-2 operators which schematically have the following form:
\begin{IEEEeqnarray}{c}
\mathcal{O}_S = Tr\left[\mathcal{Z} \, \mathcal{D}_+^S \, \mathcal{Z}\right] + \ldots
\end{IEEEeqnarray}
The case of long AdS$_3$ strings has been extensively studied in the literature because the corresponding field theory operators play a vital role in DIS. The form of their anomalous dimensions admits the following strong coupling expansion:
\small\begin{IEEEeqnarray}{ll}
\mathcal{E} - \mathcal{S} = \rho_c \, \ln\mathcal{S} &+ \sum_{n = 0}^{\infty}\sum_{k = 0}^{n} \rho_{(nk)} \, \frac{\ln^{k}\mathcal{S}}{\mathcal{S}^n} = \rho_c \, \ln\mathcal{S} + \rho_0 + \sum_{n = 1}^{\infty}\rho_{(nn)}\frac{\ln^n\mathcal{S}}{\mathcal{S}^n} + \sum_{n = 2}^{\infty}\rho_{(nn-1)}\frac{\ln^{n-1}\mathcal{S}}{\mathcal{S}^{n}} + \nonumber \\
& + \sum_{n = 3}^{\infty}\rho_{(nn-2)}\frac{\ln^{n-2}\mathcal{S}}{\mathcal{S}^{n}} + \ldots + \frac{\rho_{1}}{\mathcal{S}} + \frac{\rho_{2}}{\mathcal{S}^2} + \frac{\rho_{3}}{\mathcal{S}^3} + \ldots, \label{AnomalousDimensionsI1}
\end{IEEEeqnarray} \normalsize
where $\mathcal{S} = \pi S/ 2 \sqrt{\lambda}$. By expressing the above result in terms of Lambert's function $W_{-1}\left(-e^{-4\mathcal{S} - 3/2}/4\right)$, \\
\footnotesize\begin{IEEEeqnarray}{ll}
\gamma = & - \frac{W_{-1}}{2} - \left(2\,\mathcal{S} + \frac{5}{4}\right) + \frac{9}{8\,W_{-1}} - \bigg[\frac{\mathcal{S}}{2} + \frac{35}{16}\bigg]\frac{1}{\left(W_{-1}\right)^2} + \bigg[\frac{5\,\mathcal{S}}{2} + \frac{2213}{384}\bigg]\frac{1}{\left(W_{-1}\right)^3} - \bigg[\mathcal{S}^2 + \frac{361\,\mathcal{S}}{32} + \frac{6665}{384}\bigg]\cdot \nonumber \\[6pt]
& \cdot\frac{1}{\left(W_{-1}\right)^4} + \bigg[\frac{19\,\mathcal{S}^2}{2} + \frac{1579\,\mathcal{S}}{32} + \frac{433501}{7680}\bigg]\frac{1}{\left(W_{-1}\right)^5} - \bigg[\frac{10\,\mathcal{S}^3}{3} + \frac{259\,\mathcal{S}^2}{4} + \frac{81.799\,\mathcal{S}}{384} + \frac{2.963.887}{15.360}\bigg]\frac{1}{\left(W_{-1}\right)^6} + \nonumber \\[6pt]
& + \bigg[\frac{136\,\mathcal{S}^3}{3} + \frac{3069\,\mathcal{S}^2}{8} + \frac{175.481\,\mathcal{S}}{192} + \frac{2.350.780.111}{3.440.640}\bigg]\frac{1}{\left(W_{-1}\right)^7} - \ldots, \label{AnomalousDimensionsI2}
\end{IEEEeqnarray} \normalsize \\
we have analytically calculated the following coefficients:
\begin{IEEEeqnarray}{l}
\rho_{c} = \frac{1}{2} \quad , \quad \rho_0 = 2\ln2 - \frac{1}{2} \quad , \quad \rho_{1} = \frac{\ln 2}{2} - \frac{1}{8} \quad , \quad \rho_{2} = -\frac{\ln^2 2}{4} + \frac{9\ln2}{32} - \frac{5}{128} \nonumber
\end{IEEEeqnarray}
and also the following three infinite series of coefficients:
\begin{IEEEeqnarray}{ll}
\rho_{(mm)} = \frac{\left(-1\right)^{m + 1}}{4^m} \frac{1}{2m}, \nonumber \\[12pt]
\rho_{(m+1,m)} = \frac{1}{2} \frac{\left(-1\right)^{m + 1}}{4^{m + 1}} \Big[H_m + \frac{m}{4} +& 1 - 4\ln2\Big] \nonumber \\[12pt]
\rho_{(m+2,m)} = \frac{\left(-1\right)^{m + 1}}{4^{m+3}} \cdot \left(m + 1\right) \cdot \Bigg\{&H^2_{m + 1} - H^{(2)}_{m + 1} + \frac{1}{2} \left(m - 16\ln 2 + 5\right) \cdot H_{m + 1} + \frac{m^2}{24} - \nonumber \\
& - \left(2\ln 2 + \frac{1}{24}\right) m + 16\ln^2 2 - 10 \ln2\Bigg\}. \nonumber
\end{IEEEeqnarray}
The series $\rho_{(mm)}$ and $\rho_{(m+1,m)}$ were derived for the first time in \cite{GeorgiouSavvidy11}. The infinite series of the next-to-next-to-leading coefficients $\rho_{(m+2,m)}$ is derived in section \ref{SpinFunctionI}. \\
\section[Gubser-Klebanov-Polyakov $\mathbb{R}\times\text{S}^2$ String]{Gubser-Klebanov-Polyakov $\mathbb{R}\times\text{S}^2$ String\label{GKPStringII}}
In \cite{GubserKlebanovPolyakov02} the following configuration of a string that has its center at the north pole of $\text{S}^2 \subset \text{S}^5$ and rotates around it was considered:
\begin{IEEEeqnarray}{c}
\Big\{t = \kappa \tau, \, \rho = \theta = \phi_1 = \phi_2 = 0\Big\} \times \Big\{\overline{\theta}_1 = \overline{\theta}_1\left(\sigma\right), \, \overline{\theta}_2 = \kappa \omega \tau, \, \overline{\phi}_1 = \overline{\phi}_2 = \overline{\phi}_3 = 0\Big\}, \qquad \label{GKPAnsatzII1}
\end{IEEEeqnarray}
where the line element of AdS$_5 \times \text{S}^5$ is
\begin{IEEEeqnarray}{ll}
ds^2 = \ell^2 \Big[-\cosh^2\rho & dt^2 + d\rho^2 + \sinh^2\rho \, \Big(d\theta^2 + \cos^2\theta \, d\phi_1^2 + \sin^2\theta \, d\phi_2^2\Big) + \nonumber \\[6pt]
& + d\overline{\theta}_1^2 + \cos^2\overline{\theta}_1 \, d\overline{\phi}_1^2 + \sin^2\overline{\theta}_1 \, \left(d\overline{\theta}_2^2 + \cos^2\overline{\theta}_2 \, d\overline{\phi}_2^2 + \sin^2\overline{\theta_2} \, d\overline{\phi}_3^2\right)\Big]. \qquad
\end{IEEEeqnarray}
\indent We shall now briefly demonstrate that the above GKP solution is unique in the sense that it minimizes the energy of an $\mathbb{R}\times\text{S}^2$ string with a single, constant angular momentum $J$. Let us consider the following generic ansatz on $\mathbb{R} \times \text{S}^2$:
\begin{IEEEeqnarray}{c}
\Big\{t = \kappa \tau, \, \rho = \theta = \phi_1 = \phi_2 = 0\Big\} \times \Big\{\overline{\theta}_1 = \overline{\theta}_1\left(\tau,\sigma\right), \, \overline{\theta}_2 = \overline{\theta}_2\left(\tau,\sigma\right), \, \overline{\phi}_1 = \overline{\phi}_2 = \overline{\phi}_3 = 0\Big\}, \qquad \label{GKPAnsatzII2}
\end{IEEEeqnarray}
which leads to the Polyakov action (in the conformal gauge, $\gamma_{ab} = \eta_{ab}$),
\begin{IEEEeqnarray}{c}
\mathcal{S}_P = \frac{\ell^2}{4\pi \alpha'} \int \left(-\dot{t}^2 + \dot{\overline{\theta}}_1^2 - \overline{\theta}_1'^2 + \sin^2\overline{\theta}_1 \big(\dot{\overline{\theta}}_2^2 - \overline{\theta}_2'^2\big)\right) d\tau d\sigma.
\end{IEEEeqnarray}
One can derive the Hamiltonian density of the string in the standard way, namely
\begin{IEEEeqnarray}{c}
\mathcal{H} = \frac{\ell^2}{4\pi \alpha'} \left\{- \kappa^2 + \frac{p_1^2}{4} + \overline{\theta}_1'^2 +
\frac{p_2^2}{4\sin^2 \overline{\theta}_1} + \overline{\theta}_2'^2 \, \sin^2\overline{\theta}_1\right\} = 0\,, \quad \begin{array}{l} p_1 = 2\,\dot{\overline{\theta}}_1 \\ p_2 = 2\,\dot{\overline{\theta}}_2\,\sin^2\overline{\theta}_1.\end{array} \qquad \label{VirasoroII1}
\end{IEEEeqnarray}
\eqref{VirasoroII1} is identical to one of the Virasoro constraints. We want to find the minimum of the energy $E$ under the constraint of fixed angular momentum $J$. To achieve this we introduce the Lagrange multiplier $\omega$ and look for the minimum of the quantity
\begin{IEEEeqnarray}{c}
Y \equiv \frac{\ell^2}{4\pi \alpha'} \int d\sigma \cdot 2 \, \sqrt{\frac{p_1^2}{4} + \overline{\theta}_1'^2 +
\frac{p_2^2}{4\sin^2\overline{\theta}_1} + \overline{\theta}_2'^2\,\sin^2\overline{\theta}_1} - \omega \left[\frac{\ell^2}{4\pi\alpha'} \int p_2 \, d\sigma -J\right], \qquad
\end{IEEEeqnarray}
which is obtained after substituting $\kappa$ from \eqref{VirasoroII1} into the integral of the energy, $E = \ell^2/4\pi \alpha'\int d\sigma \cdot 2 \kappa$. We find:
\begin{IEEEeqnarray}{c}
\frac{\delta Y}{\delta p_1} = 0 \Rightarrow p_1 = 0 \Rightarrow \dot{\overline{\theta}}_1 = 0 \Rightarrow \overline{\theta}_1 = \overline{\theta}_1\left(\sigma\right) \label{GKPMinimumII1} \\[6pt]
\frac{\delta Y}{\delta p_2} = 0 \Rightarrow \dot{\overline{\theta}}_2 = \omega \, \kappa \Rightarrow \overline{\theta}_2 = \kappa \, \omega \, \tau + \widetilde{\phi}\left(\sigma\right).
\end{IEEEeqnarray}
\begin{figure}
\begin{center}
\includegraphics[scale=0.4]{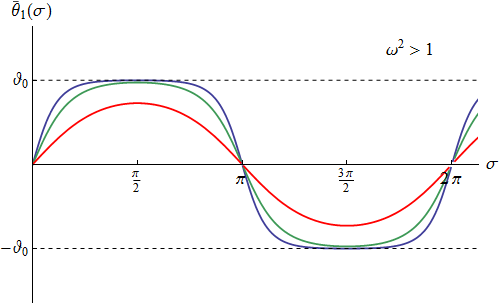}
\qquad
\includegraphics[scale=0.4]{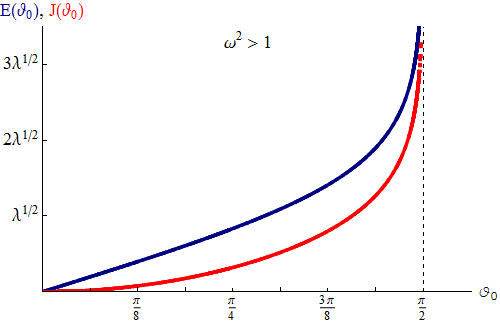}
\caption{$\overline{\theta} = \overline{\theta}_1\left(\sigma\right)$ and energy/spin of the folded closed $\mathbb{R}\times\text{S}^2$ string \eqref{GKPAnsatzII1} for $\omega > 1$.} \label{Graph:GKPIIa}
\end{center}
\end{figure}
\indent One can now use the second Virasoro constraint $\dot{\overline{\theta}}_1\,\overline{\theta}_1' + \dot{\overline{\theta}}_2 \, \overline{\theta}_2' \, \sin^2\overline{\theta}_1 = 0$ and equation \eqref{GKPMinimumII1} to conclude that $\overline{\theta}_2' = 0$ or equivalently $\widetilde{\phi}\left(\sigma\right)=0$. Let us mention that the vanishing of the functional derivative $\delta Y / \delta\overline{\theta}_1(\sigma) = 0$ will give the equation of motion for $\overline{\theta}_1$. We thus end up with the GKP solution \eqref{GKPAnsatzII1}, the Polyakov action of which is given by:
\small\begin{IEEEeqnarray}{ll}
\mathcal{S}_P & = \frac{\ell^2}{4 \pi \alpha'} \int \left(-\dot{t}^2 + \dot{\overline{\theta}}_2^2\sin^2\overline{\theta}_1 -\overline{\theta}_1'^2\right) d\tau d\sigma = \frac{\ell^2}{4 \pi \alpha'} \int \left(-\kappa^2 + \kappa^2 \omega^2 \sin^2\overline{\theta}_1 -\overline{\theta}_1'^2\right) d\tau d\sigma. \qquad \label{GKPActionII}
\end{IEEEeqnarray} \normalsize
\indent Setting $\sigma\left(\vartheta_0\right) = \pi/2$ for the length of the string, we obtain the following cases, depending on the value of $\omega \neq 1$: \\[6pt]
\begin{figure}
\begin{center}
\includegraphics[scale=0.4]{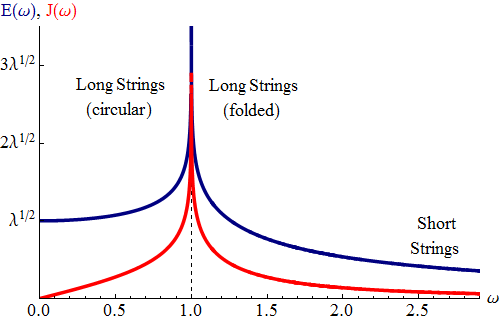}
\qquad
\includegraphics[scale=0.4]{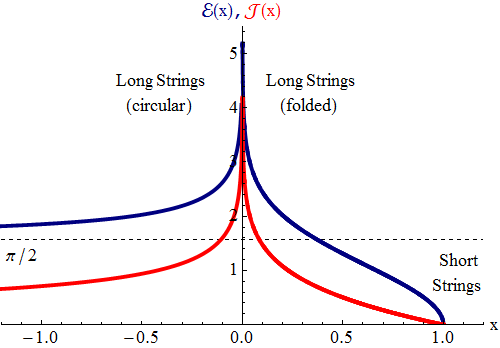}
\caption{Energy/spin of the closed folded/circular $\mathbb{R}\times\text{S}^2$ string as functions of $\omega$ and $x$.} \label{Graph:GKPIIb}
\end{center}
\end{figure}
$\underline{\omega^2 < 1}$. Circular folded string.
\begin{IEEEeqnarray}{l}
\overline{\theta}_1\left(\sigma\right) = am \left[\kappa \, \sigma \, \Big| \, \omega^2\right] \quad , \quad \kappa = \frac{2}{\pi} \cdot \mathbb{K}\left(\omega^2\right) \label{GKPIICircularLength1}
\end{IEEEeqnarray}
\begin{IEEEeqnarray}{c}
E\left(\omega\right) = \frac{2\sqrt{\lambda}}{\pi} \cdot \mathbb{K}\left(\omega^2\right) \Rightarrow \mathcal{E} \equiv \frac{\pi \, E}{2 \, \sqrt{\lambda}} = \mathbb{K}\left(1 - \widetilde{x}\right) \label{GKPIICircularEnergy1}
\end{IEEEeqnarray}
\begin{IEEEeqnarray}{c}
J\left(\omega\right) = \frac{2\sqrt{\lambda}}{\pi\,\omega} \Big[\mathbb{K}\left(\omega^2\right) - \mathbb{E}\left(\omega^2\right)\Big] \Rightarrow \mathcal{J} \equiv \frac{\pi \, J}{2 \, \sqrt{\lambda}} = \frac{1}{\sqrt{1-\widetilde{x}}} \Big[\mathbb{K}\left(1 - \widetilde{x}\right) - \mathbb{E}\left(1 - \widetilde{x}\right)\Big]. \qquad \label{GKPIICircularSpin1}
\end{IEEEeqnarray} \\
$\underline{\omega^2 > 1}$. Folded closed string.
\begin{IEEEeqnarray}{l}
\overline{\theta}_1\left(\sigma\right) = am \left[\kappa \, \sigma \, \Big| \, \omega^2\right], \quad \kappa = \frac{2}{\pi\,\omega} \cdot \mathbb{K}\left(\frac{1}{\omega^2}\right), \quad \omega = \csc \vartheta_0 \label{GKPIILength3}
\end{IEEEeqnarray}
\begin{IEEEeqnarray}{l}
E\left(\omega\right) = \frac{2\sqrt{\lambda}}{\pi\,\omega} \cdot \mathbb{K}\left(\frac{1}{\omega^2}\right) \Rightarrow \mathcal{E} \equiv \frac{\pi \, E}{2 \, \sqrt{\lambda}} = \sqrt{1 - x} \cdot \mathbb{K}\left(1 - x\right) \label{GKPIIEnergy2}
\end{IEEEeqnarray}
\begin{IEEEeqnarray}{l}
J\left(\omega\right) = \frac{2\sqrt{\lambda}}{\pi} \cdot \left[\mathbb{K}\left(\frac{1}{\omega^2}\right) - \mathbb{E}\left(\frac{1}{\omega^2}\right)\right] \Rightarrow \mathcal{J} \equiv \frac{\pi \, J}{2 \, \sqrt{\lambda}} = \mathbb{K}\left(1 - x\right) - \mathbb{E}\left(1 - x\right) \qquad \label{GKPIISpin2}
\end{IEEEeqnarray}
\begin{IEEEeqnarray}{l}
\gamma \equiv \mathcal{E} - \mathcal{J} = \left(\sqrt{1 - x} - 1\right) \cdot \mathbb{K}\left(1 - x\right) + \mathbb{E}\left(1 - x\right).
\end{IEEEeqnarray} \\
$x \equiv 1 - 1/\omega^2$ and $\widetilde{x} \equiv 1 - \omega^2$ are the complementary parameters of $1/\omega^2$ and $\omega^2$ respectively. In figures \ref{Graph:GKPIIa}--\ref{Graph:GKPIIb} we have plotted $\overline{\theta}_1 = \overline{\theta}_1\left(\sigma\right)$ for various values of $\omega > 1$ as well as the energy and the spin of the $\mathbb{R}\times\text{S}^2$ string as functions of $\vartheta_0$, $\omega$ and $x$. \\
\indent For $\omega > 1$, there exist two interesting regimes where one would want to obtain the functional dependence of $E = E\left(J\right)$ and the corresponding anomalous dimensions $\gamma = \gamma\left(J\right)$, namely the short-string limit $\omega \rightarrow \infty$ and the long-string limit $\omega \rightarrow 1^{+}$. In what follows we shall be concerned only with the latter. Short strings are briefly treated in appendix \ref{Short-Slow-StringsAppendix}. \\
\subsection[Long Folded Strings]{Long Folded Strings in $\mathbb{R}\times\text{S}^2$: $\omega \rightarrow 1^+$, $J \gg \sqrt{\lambda}$ \label{LongStringsII}}
For long folded strings on $\text{S}^2$ ($\omega \rightarrow 1^{+}$), the expansions for the energy and the spin become (cf. appendix \ref{EllipticFunctionsAppendix}):
\small\begin{IEEEeqnarray}{lll}
E & = \frac{\sqrt{\lambda}}{\pi^2\omega} \cdot \sum_{n = 0}^\infty & \left(\frac{\Gamma\left(n + 1/2\right)}{n!}\right)^2 \Big[2\psi\left(n + 1\right) - 2\psi\left(n + 1/2\right) - \ln\left(1 - 1 / \omega^2\right)\Big] \cdot \left(1 - 1 / \omega^2\right)^n \qquad \\[12pt]
J & = \frac{\sqrt{\lambda}}{\pi} \cdot \Bigg\{4&\ln2 - 2 - \ln\left(1 - 1 / \omega^2\right) - \frac{1}{2\pi} \sum_{n = 0}^\infty \frac{\Gamma\left(n + 1/2\right)\Gamma\left(n + 3/2\right)}{\left(\left(n + 1\right)!\right)^2} \, \Big[2 \, \psi\left(n + 1\right) - \nonumber \\[6pt]
&& - 2 \, \psi\left(n + 1/2\right) - \ln\left(1 - 1 / \omega^2\right) + \frac{2n}{\left(n + 1\right)\left(2n + 1\right)}\Big] \cdot \left(1 - 1 / \omega^2\right)^{n + 1} \Bigg\}. \qquad
\end{IEEEeqnarray} \normalsize
In terms of the complementary parameter $x \equiv 1 - 1 / \omega^2 \rightarrow 0^+$ the above series can be written in compact forms as follows:
\small \begin{IEEEeqnarray}{c}
\mathcal{E} \equiv \frac{\pi\,E}{2\sqrt{\lambda}} = \sqrt{1 - x} \cdot \sum_{n = 0}^{\infty} x^n\left(d_n\ln x + h_n\right) = - \sum_{n = 0}^{\infty} x^n \cdot \sum_{k = 0}^n \frac{(2k - 3)!!}{\left(2k\right)!!} \left(d_{n - k} \ln x + h_{n - k}\right) \qquad \label{GKPIIEnergy3} \\[6pt]
\mathcal{J} \equiv \frac{\pi\,J}{2\sqrt{\lambda}} = \sum_{n = 0}^{\infty} x^n\left(c_n\ln x + b_n\right). \label{GKPIISpin3}
\end{IEEEeqnarray} \normalsize
The coefficients that appear in series \eqref{GKPIIEnergy3} and \eqref{GKPIISpin3} are given by:\footnote{Some useful values of the double factorial are: $0!! = 1$, $\left(-1\right)!! = 1$, $\left(-3\right)!! = -1$.}
\begin{IEEEeqnarray}{c}
d_n = - \frac{1}{2} \left(\frac{\left(2n - 1\right)!!}{\left(2n\right)!!}\right)^2 \,, \qquad h_n = - 4\,d_n \cdot \left(\ln2 + H_n - H_{2n}\right) \nonumber \\
c_n = - \frac{d_n}{2n - 1}\,, \qquad b_n = - 4\,c_n \cdot \left[\ln2 + H_n - H_{2n} + \frac{1}{2\left(2n - 1\right)}\right], \label{LongSeriesCoefficientsII1}
\end{IEEEeqnarray}
for $n = 0 \,, 1 \,, 2 \,, \ldots$ \\
\subsection[Short-Long Strings Duality]{Short-Long Strings Duality \label{ShortLongDuality}}
\begin{figure}
\begin{center}
\includegraphics[scale=0.4]{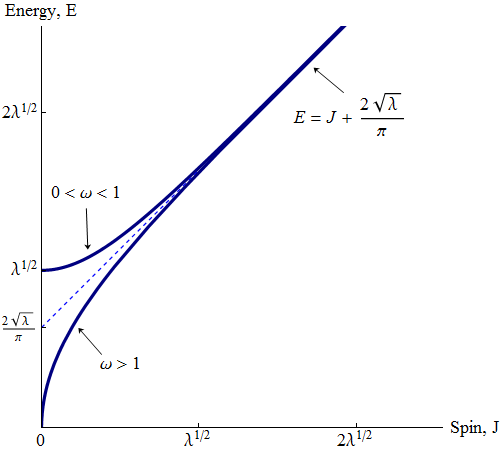}
\qquad
\includegraphics[scale=0.4]{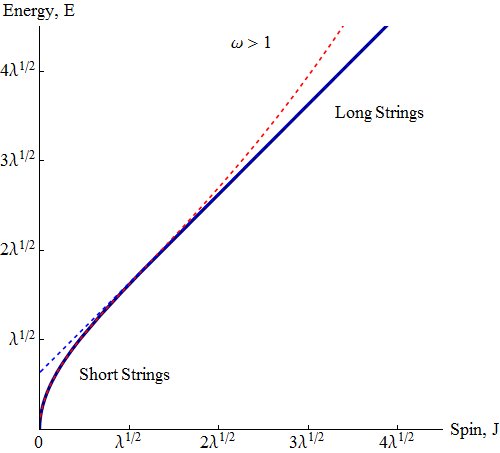}
\caption{$E = E\left(J\right)$ of the folded closed $\mathbb{R}\times\text{S}^2$ string.} \label{Graph:Energy-SpinGKPII}
\end{center}
\end{figure}
Following \cite{GeorgiouSavvidy11}, we may now write down a formula that connects the values of conserved charges at the two opposite ends of the closed folded string spectrum, that is "short" strings ($\omega \rightarrow \infty$) and "long" strings ($\omega \rightarrow 1^+$). There's a known expression between complete elliptic integrals of the first and second kind, namely Legendre's relation (see e.g. \cite{AbramowitzStegun65}):
\begin{IEEEeqnarray}{c}
\mathbb{E}(k) \mathbb{K}(k') + \mathbb{K}(k) \mathbb{E}(k') - \mathbb{K}(k) \mathbb{K}(k') = \frac{\pi}{2}, \label{Legendre}
\end{IEEEeqnarray}
where the arguments of elliptic integrals, $k = 1 / \omega^2$ and $k' = x = 1 / \omega' \, ^2$ satisfy $k + k' = 1$. Solving \eqref{GKPIIEnergy2}--\eqref{GKPIISpin2} for $\mathbb{E}(k)$ and $\mathbb{K}(k)$ and substituting in \eqref{Legendre}, we get the following duality relation between classical folded short and long strings that spin on $\text{S}^2 \subset \text{S}^5$:
\begin{IEEEeqnarray}{c}
\omega \, \omega' \, E E' - \omega \, E J' - \omega' \, E' J = \frac{2 \lambda}{\pi},\, \qquad \omega>1. \label{Short-LongII1}
\end{IEEEeqnarray}
\indent This relation is completely analogous to the one found for closed folded strings that spin inside AdS$_3$, \eqref{Short-Long2a}. One could also write \eqref{Short-LongII1} in terms of $\gamma \equiv E - S$. Plotting the functions $E = E\left(\omega\right)$ and $J = J\left(\omega\right)$ in a common diagram parametrically, the graph of $E = E\left(J\right)$ along with the first 4 terms of its "short" series \eqref{ShortStringIIEnergy3} and the first 2 terms of its "long" approximation \eqref{MathematicaAnomalousDimensionsII1} has been obtained in figure \ref{Graph:Energy-SpinGKPII} (red and blue dashed lines respectively of the plot on the right). Likewise, a similar relation can be formulated for fast and slow circular strings using \eqref{GKPIICircularEnergy1}--\eqref{GKPIICircularSpin1}:
\begin{IEEEeqnarray}{c}
E E' - \omega' \, E J' - \omega \, E' J = \frac{2 \lambda}{\pi},\, \qquad \omega<1, \label{Short-LongII2}
\end{IEEEeqnarray}
where $\widetilde{k} = \omega^2$, $\widetilde{k}' = \widetilde{x} = \omega'^2$ and $\widetilde{k} + \widetilde{k}' = 1$. \\
\section[Inverse Spin Functions and Anomalous Dimensions on $\mathbb{R}\times\text{S}^2$]{Inverse Spin Functions and Anomalous Dimensions on $\mathbb{R}\times\text{S}^2$\label{SpinFunctionII}}
\subsection[Inverse Spin Function]{Inverse Spin Function \label{InverseSpinFunctionII}}
We will now follow the method of \cite{GeorgiouSavvidy11} in order to invert the J-series \eqref{GKPIISpin3} for $x = x\left(\mathcal{J}\right)$ and obtain $\mathcal{E} = \mathcal{E}\left(\mathcal{J}\right)$ by substituting $x\left(\mathcal{J}\right)$ into $\mathcal{E}\left(x\right)$. Let us first solve \eqref{GKPIISpin3} for $\ln x$:
\begin{IEEEeqnarray}{ll}
\mathcal{J} = \sum_{n = 0}^{\infty} x^n\Big(c_n\ln x + b_n\Big) &\Rightarrow \ln x = \left[\frac{\mathcal{J} - b_0}{c_0} - \sum_{n = 1}^{\infty} \frac{b_n}{c_0} \, x^n\right] \cdot \sum_{n = 0}^{\infty} \left(-1\right)^n \left(\sum_{k = 1}^{\infty} \frac{c_k}{c_0} \, x^k\right)^{n}. \qquad \label{InverseSpinEquationII}
\end{IEEEeqnarray}
Performing the products between the series and exponentiating, we are led to the following equation that we will eventually have to solve (or invert) for $x$:
\begin{IEEEeqnarray}{c}
x = x_0 \cdot \exp\left[\sum_{n = 1}^{\infty} \text{a}_n \, x^n\right] = x_0 \cdot \exp\left(\text{a}_1 \, x + \text{a}_2 \, x^2 + \text{a}_3 \, x^3 + \ldots\right), \label{xEquationII1}
\end{IEEEeqnarray}
where
\begin{IEEEeqnarray}{c}
x_0 \equiv \exp\left[\frac{\mathcal{J} - b_0}{c_0}\right] = 16 \, e^{-2\mathcal{J} - 2} \label{DefinitionIIx_0}
\end{IEEEeqnarray}
solves \eqref{InverseSpinEquationII} to lowest order in $x$ and the coefficients a$_n$ are determined from \eqref{InverseSpinEquationII}. One possible way to revert series \eqref{xEquationII1} with respect to the variable $x$, is by means of the Lagrange inversion theorem \cite{AbramowitzStegun65}. In our case it turns out that the function to be inverted has a special form that significantly simplifies the computation of its inverse. The following reversion formula (applied here to the exponential function) is named after J.-L. Lagrange and H. H. B\"{u}rmann \cite{WhittakerWatson27}:
\begin{IEEEeqnarray}{c}
x = \sum_{n = 1}^{\infty} \frac{x_0^n}{n!} \cdot \left\{\frac{d^{n - 1}}{dz^{n - 1}} \exp\left[\sum_{m = 1}^{\infty} n \, \text{a}_m \, z^m\right]\right\}_{z = 0}.
\end{IEEEeqnarray}
We find
\begin{IEEEeqnarray}{ll}
x = \sum_{n = 1}^{\infty} x_0^n \cdot \sum_{k,j_i = 0}^{n - 1} \frac{n^k}{n!} \, {n - 1 \choose j_1 \,,\; j_2 \,,\; \ldots \,,\; j_{n - 1}} \, \text{a}_1^{j_1} \text{a}_2^{j_2} \ldots \text{a}_{n - 1}^{j_{n - 1}}, \qquad \label{xEquationII2}
\end{IEEEeqnarray}
where
\begin{IEEEeqnarray}{l}
j_1 + j_2 + \ldots + j_{n - 1} = k \quad \& \quad j_1 + 2 \, j_2 + \ldots + \left(n - 1\right) j_{n - 1} = n - 1. \nonumber
\end{IEEEeqnarray}
\indent Now notice from \eqref{InverseSpinEquationII} that the $\text{a}_i$'s can only be linear functions of $\mathcal{J}$, so that the inverse spin function $x = x\left(\mathcal{J}\right)$ has to be of the form:
\begin{IEEEeqnarray}{c}
x = \sum_{n = 1}^{\infty} x_0^n \cdot \sum_{k = 0}^{n - 1} a_{nk} J^k, \label{xEquationII3}
\end{IEEEeqnarray}
where $a_{nk}$ are some constants. This follows from the two constraints on the values of $j$ which imply
\begin{IEEEeqnarray}{c}
\begin{array}{c} j_1 + j_2 + \ldots + j_{n - 1} = k \\[6pt] j_1 + 2 \, j_2 + \ldots + \left(n - 1\right) j_{n - 1} = n - 1 \end{array}\Bigg\} \Rightarrow k + j_2 + \ldots + \left(n - 2\right) j_{n - 1} = n - 1, \qquad
\end{IEEEeqnarray}
that is the power of $J$ is at most $k \leq n - 1$. A second conclusion that can be drawn from these two constraints is that all leading in $\mathcal{J}$ contributions to $x$ are controlled by the leading in $\mathcal{J}$ terms of $\text{a}_1$, all subleading $\mathcal{J}$-contributions to $x$ are controlled by $\text{a}_1$ and the leading in $\mathcal{J}$ terms of $\text{a}_2$, etc., i.e.\ $x\left(\mathcal{J}\right)$ has all its coefficients up to $x_0^n\,\mathcal{J}^{n-m}$ controlled by $\text{a}_1 \ , \ldots \ \text{a}_{m - 1}$, and the leading term of $\text{a}_m$. To see this, notice from $k + j_2 + \ldots + \left(n - 2\right) j_{n - 1} = n - 1$ that when some $j_m$ in \eqref{xEquationII2} is $j_m \neq 0$ (minimum 1), $k = j_m + \ldots + j_{n - 1}$ is at most $n - 1 - \left(m - 1\right) = n - m$. This is consistent with our expectations from \eqref{xEquationII1}, from which the same conclusion about the number of terms that should be kept on the r.h.s., in order to fully determine $x\left(\mathcal{J}\right)$ up to a given order, is reached. \\
\subsection[Anomalous Dimensions]{Anomalous Dimensions \label{AnomalousDimensionsII}}
Having derived a general formula for $x(\mathcal{J})$, we may express the anomalous scaling dimensions $\gamma = \mathcal{E} - \mathcal{J}$ of the $\mathbb{R}\times\text{S}^2$ spinning closed and folded string as a function of $\mathcal{J}$:
\begin{IEEEeqnarray}{l}
\mathcal{E} - \mathcal{J} = \sum_{n = 0}^{\infty} x^n \left(f_n \, \ln x + g_n\right) = \sum_{n = 0}^{\infty} x^n \left[A_n + f_n \, \ln \frac{x}{x_0}\right]\,, \quad \mathcal{E} \equiv \frac{\pi E}{2\sqrt{\lambda}}\,, \quad \mathcal{J} \equiv \frac{\pi J}{2\sqrt{\lambda}}, \qquad \label{AnomalousDimensions4}
\end{IEEEeqnarray}
where
\begin{IEEEeqnarray}{l}
f_n \equiv -c_n - \sum_{k = 0}^n \frac{(2k - 3)!!}{\left(2k\right)!!} \cdot d_{n - k} \,, \ g_n \equiv -b_n - \sum_{k = 0}^n \frac{(2k - 3)!!}{\left(2k\right)!!} \cdot h_{n - k} \,, \ n = 0,1,2,\ldots \qquad
\end{IEEEeqnarray}
and the coefficients $A_n$ are defined as:
\begin{IEEEeqnarray}{ll}
A_n \equiv g_n + f_n \ln x_0 = g_n + 2 f_n \, \left(2\ln 2 - \mathcal{J} - 1\right).
\end{IEEEeqnarray}
\indent For large $\mathcal{J}$, series \eqref{InverseSpinEquationII} may be inverted for $x = x\left(\mathcal{J}\right)$ using a computer algebra system. The inverse spin function $x\left(\mathcal{J}\right)$ may then be plugged into equation \eqref{AnomalousDimensions4} and give the anomalous dimensions of $\mathcal{N} = 4$ SYM operators $\text{Tr}\left[\Phi\mathcal{Z}^m \, \Phi \, \mathcal{Z}^{J-m}\right] + \ldots$ in terms of the (large) R-charge $\mathcal{J}$. Such a calculation has been performed with Mathematica and both series of the inverse spin function $x = x\left(\mathcal{J}\right)$ and anomalous dimensions $\gamma = \gamma\left(\mathcal{J}\right)$ have been obtained (cf. \eqref{Exact_Value_of_xII1}, \eqref{MathematicaAnomalousDimensionsII1}). It turns out that both expansions contain the following terms:
\begin{IEEEeqnarray}{l}
\text{Leading Terms (L): } \mathcal{J}^{n-1} \, \left(e^{-2\mathcal{J} - 2}\right)^n \nonumber \\
\text{Next-to-Leading/Subleading Terms (NL): } \mathcal{J}^{n-2} \, \left(e^{-2\mathcal{J} - 2}\right)^n \nonumber \\
\text{NNL Terms: } \mathcal{J}^{n-3} \, \left(e^{-2\mathcal{J} - 2}\right)^n \nonumber \\
\hspace{2cm} \vdots
\end{IEEEeqnarray}
Note that series \eqref{GKPIIEnergy3}, \eqref{GKPIISpin3} and \eqref{AnomalousDimensions4} are identical in structure. We may then prove that in order to get $\mathcal{E} - \mathcal{J}$ up to a given subleading order, the inverse spin function $x\left(\mathcal{J}\right)$ that will be inserted into \eqref{AnomalousDimensions4} has to be known up to no more than the same order. From \eqref{xEquationII1} and \eqref{xEquationII3} we write:
\begin{IEEEeqnarray}{c}
\ln \frac{x}{x_0} = \sum_{k = 1}^{\infty} \text{a}_k \, x^k = \text{a}_1 \, x + \text{a}_2 \, x^2 + \text{a}_3 \, x^3 + \ldots \nonumber \\
x = \sum_{n = 1}^{\infty} x_0^n \cdot \sum_{k = 0}^{n - 1} a_{nk} \mathcal{J}^k = \sum_{n = 1}^{\infty} \mathcal{J}^{n - 1} \, x_0^n \cdot \sum_{k = 0}^{n - 1} \frac{\widetilde{a}_{nk}}{\mathcal{J}^k} = \frac{1}{\mathcal{J}} \sum_{n = 1}^{\infty} \mathcal{J}^n \, x_0^n \cdot \sum_{k = 0}^{n - 1} \frac{\widetilde{a}_{nk}}{\mathcal{J}^k}, \qquad \label{xEquationII4}
\end{IEEEeqnarray}
where $a_{nk} = \widetilde{a}_{n(n - k - 1)}$ are some constants and $\text{a}_n$ are linear functions of $\mathcal{J}$. The last equation follows from \eqref{xEquationII3} after some reshuffling. The anomalous dimensions \eqref{AnomalousDimensions4} are then written as follows:
\small \begin{IEEEeqnarray}{ll}
\mathcal{E} - \mathcal{J} & = \sum_{n = 0}^{\infty} x^n \left(f_n \, \ln x + g_n\right) = \sum_{n = 0}^{\infty} x^n \left[A_n + f_n \, \ln \frac{x}{x_0}\right] = \sum_{n = 0}^{\infty} x^n \left[A_n + \sum_{k = 1}^{\infty} f_n \, \text{a}_k \, x^k\right]. \qquad \label{AnomalousDimensions6}
\end{IEEEeqnarray} \normalsize
\indent Since all the leading terms of $x^n$ are of the order $1/\mathcal{J}^n$ and they multiply either $A_n$ or $f_{n-k}\cdot\text{a}_k$ in \eqref{AnomalousDimensions6}, which are both linear in $\mathcal{J}$, we see that the $r$-th subleading term of $\mathcal{E} - \mathcal{J}$ (which is of the order $1/\mathcal{J}^r$) cannot get any contributions from its $x^{r + 2}$ terms. Therefore in order to get precisely the first $r$-subleading orders of the anomalous dimensions $\mathcal{E} - \mathcal{J}$ ($r = 1$ leading, $r = 2$ subleading, etc.), no more than the first $r + 1$ powers of $x$ need to be retained in \eqref{AnomalousDimensions6}. In addition, the last power of $x$ to be kept in \eqref{AnomalousDimensions6} (namely $x^{r + 1}$) does not have to be multiplied by coefficients that do not contain $\mathcal{J}$. \\
\indent Furthermore, we can see why we need exactly $n$ subleading terms in the expansion of $x$ in order to be able to calculate $\mathcal{E} - \mathcal{J}$ up to the same subleading order $n$. Keeping less powers inside $x$ would mean that $x \cdot A_1 = - x / 4$ in \eqref{AnomalousDimensions6} essentially misses some of the subleading terms, while terms deeper than $1 / \mathcal{J}^n$ into $x$ do not contribute, since there exist no corresponding powers of $\mathcal{J}$ in the expression for $\mathcal{E} - \mathcal{J}$ that can lift them up to the wanted order. All of these observations will become clearer in what follows. \\
\subsection[Leading Terms]{Leading Terms \label{LeadingII}}
As a first application of the above, we may calculate the anomalous dimensions to leading order in $\mathcal{J}$, i.e.\ the coefficients of the following series:
\begin{IEEEeqnarray}{ll}
E - J\Big|_{(\text{L})} = \sum_{n = 1}^{\infty} \mathfrak{a}_n \, \mathcal{J}^{n - 1} \, \left(e^{-2\mathcal{J} - 2}\right)^{n}.
\end{IEEEeqnarray}
In order to be able to do this, $x$ has to be determined up to leading order in $\mathcal{J}$, i.e.\ the coefficients of the series
\begin{IEEEeqnarray}{l}
x_{(\text{\tiny{L}})} = \sum_{n = 1}^{\infty} \alpha_n \, \mathcal{J}^{n - 1} \, \left(e^{-2\mathcal{J} - 2}\right)^{n}
\end{IEEEeqnarray}
must be computed. To this end, we must keep all the terms that multiply $x^0=1$ on the r.h.s.\ of equation \eqref{InverseSpinEquationII} and just the leading in $\mathcal{J}$ terms that multiply $x^1=x$. \eqref{InverseSpinEquationII} then becomes:
\begin{IEEEeqnarray}{c}
\ln x_{(\text{\tiny{L}})} = \frac{\mathcal{J} - b_0}{c_0} - \frac{c_1}{c_0^2} \, \mathcal{J} \cdot x_{(\text{\tiny{L}})} \Rightarrow x_0 = x_{(\text{\tiny{L}})} \, \exp\left[\frac{c_1}{c_0^2} \, \mathcal{J} \cdot x_{(\text{\tiny{L}})}\right] = x_{(\text{\tiny{L}})} \, e^{\mathcal{J} \cdot x_{(\text{L})} / 2}, \qquad
\end{IEEEeqnarray}
where $x_0 = 16\,e^{-2\mathcal{J} - 2} $. This is equation \eqref{xEquationII1} for the leading terms of $x$. We either solve it by the inversion method of the previous section or we could just as well calculate the following iteration:
\begin{IEEEeqnarray}{l}
x_{(\text{\tiny{L}})} = x_0 \, e^{-x_0 \mathcal{J} / 2 \cdot e^{-x_0 \mathcal{J} / 2 \cdot e^{\ldots}}} = x_0 \cdot \tensor[^\infty]{\left(e^{- x_0 \mathcal{J} / 2}\right)}{}.
\end{IEEEeqnarray}
Given the expression for the infinite exponential (see e.g. \cite{Galidakis06}),
\begin{IEEEeqnarray}{l}
\tensor[^\infty]{\left(e^{z}\right)}{} = \frac{W\left(- z\right)}{- z} = \sum_{n = 1}^{\infty} \frac{n^{n - 1}}{n!} \, z^{n - 1},
\end{IEEEeqnarray}
where $W\left(z\right)$ is the principal branch of the Lambert W-function,\footnote{We need to choose the principal branch $W_0$ in order for $x$ to have the correct limiting behavior, i.e.\ $x \rightarrow 0^+$ as $\mathcal{J} \rightarrow +\infty$. Choosing the $W_{-1}$ branch gives $x \rightarrow -4$, to leading order. For more see also appendix \ref{LambertAppendix} that deals with the Lambert function.} we find:
\begin{IEEEeqnarray}{ll}
x_{(\text{\tiny{L}})} & = \frac{2}{\mathcal{J}} \, W\left(8 \, \mathcal{J} \, e^{-2\mathcal{J} - 2}\right) = \sum_{n = 1}^{\infty} \alpha_n \, \mathcal{J}^{n - 1} \, \left(e^{-2\mathcal{J} - 2}\right)^{n}, \qquad \label{Leading_x}
\end{IEEEeqnarray}
where we have defined
\begin{IEEEeqnarray}{ll}
\alpha_n \equiv \left(- 1\right)^{n + 1} \, 2^{3n + 1} \cdot \frac{n^{n - 1}}{n!}. \label{AlphaCoefficients}
\end{IEEEeqnarray}
\indent As we have explained above, in order to calculate the $E - J$ series to leading order in $\mathcal{J}$, we have to insert this formula for $x_{(\text{\tiny{L}})}$ into \eqref{AnomalousDimensions4} and take care as to keep only leading terms. The result is: \nocite{AbramowitzStegun65}
\begin{IEEEeqnarray}{ll}
E - J\Big|_{(\text{L})} & = \frac{2\sqrt{\lambda}}{\pi} \left\{1 +g_1 \, x_{(\text{\tiny{L}})} - 2f_2 \, \mathcal{J} x_{(\text{\tiny{L}})}^2\right\} = \frac{2\sqrt{\lambda}}{\pi} \left\{1 - \frac{x_{(\text{\tiny{L}})}}{4} - \frac{\mathcal{J} x_{(\text{\tiny{L}})}^2}{16}\right\} = \nonumber \\[12pt]
& = \frac{2\sqrt{\lambda}}{\pi} \Bigg\{1 - \frac{1}{4\mathcal{J}} \, \bigg[2 \cdot W\left(8\,\mathcal{J}\,e^{- 2\mathcal{J} - 2}\right) + W^2\left(8\,\mathcal{J}\,e^{- 2\mathcal{J} - 2}\right)\bigg]\Bigg\} = \nonumber \\[12pt]
& = \frac{2\sqrt{\lambda}}{\pi} \Bigg\{1 - \frac{1}{16} \, \sum_{n = 1}^\infty \bigg[4 \, \alpha_n + \sum_{k = 1}^{n-1} \alpha_k \, \alpha_{n - k}\bigg] \cdot \mathcal{J}^{n-1} \left(e^{-2\mathcal{J} - 2}\right)^{n} \Bigg\}, \qquad \label{AnomalousDimensions8}
\end{IEEEeqnarray}
which contains all the leading-order terms of $E - J$.\\
\subsection[Next-to-Leading Terms]{Next-to-Leading Terms \label{SubleadingII}}
We can go on and calculate all the subleading in $\mathcal{J}$ coefficients of the anomalous dimensions, i.e.\ obtain an analytic expression for the terms of the following series:
\begin{IEEEeqnarray}{ll}
E - J\Big|_{(\text{NL})} = \sum_{n = 2}^{\infty} \mathfrak{b}_n \, \mathcal{J}^{n - 2} \, \left(e^{-2\mathcal{J} - 2}\right)^{n}.
\end{IEEEeqnarray}
This time we have to know not only the leading, but also the subleading terms of $x$ in \eqref{xEquationII2}, namely
\begin{IEEEeqnarray}{ll}
x_{(\text{\tiny{NL}})} = \sum_{n = 2}^{\infty} \beta_n \, \mathcal{J}^{n - 2} \, \left(e^{-2\mathcal{J} - 2}\right)^{n}.
\end{IEEEeqnarray}
\indent This means that on the r.h.s. of \eqref{InverseSpinEquationII} we have to keep all the terms that multiply $x^{0,1}$ and only the leading in $\mathcal{J}$ terms that multiply $x^2$. Then equation \eqref{InverseSpinEquationII}, precise up to next-to-leading/subleading order, becomes:
\begin{IEEEeqnarray}{c}
\ln x_{(\text{\tiny{L}} + \text{\tiny{NL}} + \ldots)} = \frac{\mathcal{J} - b_0}{c_0} - \frac{\mathcal{J} c_1 + b_1 c_0 - b_0 c_1}{c_0^2} \cdot x_{(\text{\tiny{L}} + \text{\tiny{NL}} + \ldots)} + \frac{c_1^2 - c_0 c_2}{c_0^3} \, \mathcal{J} \cdot x_{(\text{\tiny{L}} + \text{\tiny{NL}} + \ldots)}^2 \Rightarrow \nonumber \\[6pt]
\Rightarrow x_{(\text{\tiny{L}} + \text{\tiny{NL}} + \ldots)} = x_0 \cdot \exp\left[- \frac{\mathcal{J} + 1}{2} \cdot x_{(\text{\tiny{L}} + \text{\tiny{NL}} + \ldots)} - \frac{7\mathcal{J}}{32} \cdot x_{(\text{\tiny{L}} + \text{\tiny{NL}} + \ldots)}^2\right].
\end{IEEEeqnarray}
To solve this equation, we invert it for $x_{(\text{\tiny{L}} + \text{\tiny{NL}} + \ldots)}$ by using the Lagrange-B\"{u}rmann formula, as it was done in going from equation \eqref{xEquationII1} to \eqref{xEquationII2},
\begin{IEEEeqnarray}{c}
x_{(\text{\tiny{L}} + \text{\tiny{NL}} + \ldots)} = \sum_{n = 1}^{\infty} \frac{x_0^n}{n!} \sum_{\tiny{\begin{array}{c} k,j_1 = 0 \\ n - 1 = k + j_1 \\ 0 \leq j_1 \leq k \end{array}}}^{n - 1} \left(-1\right)^k n^k \, \frac{\left(n - 1\right)!}{\left(k - j_1\right)! \, j_1!} \cdot \left(\frac{\mathcal{J} + 1}{2}\right)^{k - j_1} \left(\frac{7\mathcal{J}}{32}\right)^{j_1}. \qquad
\end{IEEEeqnarray}
\indent We now have to select and keep only the leading and next-to-leading $\mathcal{J}$-terms in this expression. Expanding the binomial in powers of $\mathcal{J}$,
\begin{IEEEeqnarray}{c}
\left(\frac{\mathcal{J} + 1}{2}\right)^{k - j_1} \cdot \left(\frac{7\mathcal{J}}{32}\right)^{j_1} = \left(\frac{1}{2}\right)^{k - j_1} \left(\frac{7\mathcal{J}}{32}\right)^{j_1} \cdot \sum_{m = 0}^{k - j_1} {k - j_1 \choose m} \, \mathcal{J}^m = \nonumber \\[6pt]
= \frac{7^{j_1}}{2^{k + 4j_1}} \cdot \sum_{m = 0}^{k - j_1} {k - j_1 \choose m} \, \mathcal{J}^{m + j_1} = \frac{7^{j_1}}{2^{k + 4j_1}} \cdot \left(\mathcal{J}^{j_1} + \ldots + \left(k - j_1\right) \mathcal{J}^{k - 1} + \mathcal{J}^k\right), \nonumber
\end{IEEEeqnarray}
we see that the leading terms $\mathcal{J}^{n - 1}\,x_0^n$, correspond to $m = k = n - 1$, $j_1 = 0$ and we obtain the leading power series \eqref{Leading_x}--\eqref{AlphaCoefficients} of the previous section. To get the next-to-leading/subleading terms $\mathcal{J}^{n - 2}\,x_0^n$, we have to put either $k = n - 1$, $j_1 = 0$, $m = n - 2$ or $k = m = n - 2$, $j_1 = 1$ and sum the two terms. We find:
\begin{IEEEeqnarray}{c}
x_{(\text{\tiny{NL}})} = \sum_{n = 1}^{\infty} \frac{x_0^n}{n!} \cdot \left\{\left(-1\right)^{n - 1} n^{n - 1} \, \frac{\left(n - 1\right)}{2^{n - 1}} + \left(-1\right)^{n - 2} n^{n - 2} \, \frac{7\left(n - 1\right)\left(n - 2\right)}{2^{n + 2}}\right\} \cdot \mathcal{J}^{n - 2}, \qquad
\end{IEEEeqnarray}
so that the leading and next-to-leading terms of $x$ are given by:
\begin{IEEEeqnarray}{ll}
x_{(\text{\tiny{L}} + \text{\tiny{NL}})} & = \sum_{n = 1}^{\infty} \left(\alpha_n \, \mathcal{J}^{n - 1} + \beta_n \, \mathcal{J}^{n - 2}\right) \cdot \left(e^{-2\mathcal{J} - 2}\right)^{n} \label{Leading+Subleading_x1}
\end{IEEEeqnarray}
with the definition ($\alpha_n$'s are defined in \eqref{AlphaCoefficients}),
\begin{IEEEeqnarray}{l}
\beta_n \equiv \left(- 1\right)^{n + 1} 2^{3n - 2} \cdot \frac{n^{n - 2}}{n!} \cdot \left(n - 1\right)\left(n + 14\right). \label{BetaCoefficients}
\end{IEEEeqnarray}
\indent Series \eqref{Leading+Subleading_x1} can be written in terms of Lambert's W-function, using formulas \eqref{Lambert1}--\eqref{Lambert6} of appendix \ref{LambertAppendix}:
\begin{IEEEeqnarray}{ll}
x_{(\text{\tiny{L}} + \text{\tiny{NL}})} & = \sum_{n = 1}^{\infty} \left(\alpha_n \, \mathcal{J}^{n - 1} + \beta_n \, \mathcal{J}^{n - 2}\right) \cdot \left(e^{-2\mathcal{J} - 2}\right)^{n} = \frac{2}{\mathcal{J}} \, W - \frac{1}{4\,\mathcal{J}^2}\frac{W^2 \left(7\,W + 8\right)}{1 + W}, \qquad \label{Leading+Subleading_x2}
\end{IEEEeqnarray}
where the argument of the W-function is $W\left(8 \, \mathcal{J} \, e^{-2\mathcal{J} - 2}\right)$. To obtain the leading and next-to-leading coefficients of $E - J$, we insert \eqref{Leading+Subleading_x2} into \eqref{AnomalousDimensions4} keeping only terms of leading and next-to-leading/subleading order:
\begin{IEEEeqnarray}{ll}
E - J\Big|_{(\text{L} + \text{NL})} & = \frac{2\sqrt{\lambda}}{\pi} \Bigg\{1 + A_1 \left(x_{(\text{\tiny{L}})} + x_{(\text{\tiny{NL}})}\right) - 2f_2 \, \mathcal{J} \, x_{(\text{\tiny{L}})}^2 - 4f_2 \, \mathcal{J} \, x_{(\text{\tiny{L}})} \cdot x_{(\text{\tiny{NL}})} + \nonumber \\[6pt]
& \hspace{1.3cm} + \left(g_2 + 2\left(2\ln 2 - 1\right)f_2\right) x_{(\text{\tiny{L}})}^2 - \left(\frac{c_1 f_2}{c_0^2} + 2f_3\right)\, \mathcal{J} \, x_{(\text{\tiny{L}})}^3 \Bigg\} = \qquad \label{AnomalousDimensions11} \\[6pt]
& = \frac{2\sqrt{\lambda}}{\pi} \Bigg\{1 - \frac{x_{(\text{\tiny{L}})}}{4} - \frac{x_{(\text{\tiny{NL}})}}{4} - \frac{\mathcal{J}}{16} \, x_{(\text{\tiny{L}})}^2 - \frac{\mathcal{J}}{8} \, x_{(\text{\tiny{L}})} \cdot x_{(\text{\tiny{NL}})} - \frac{9}{64} \, x_{(\text{\tiny{L}})}^2 - \frac{\mathcal{J}}{16} \, x_{(\text{\tiny{L}})}^3 \Bigg\}. \nonumber
\end{IEEEeqnarray}
From this expression we read off the next-to-leading/subleading coefficients as follows (the leading ones are given in \eqref{AnomalousDimensions8}):
\small\begin{IEEEeqnarray}{rlll}
&E - J\Big|_{(\text{NL})} &= - \frac{2\sqrt{\lambda}}{\pi} \Bigg\{\frac{x_{(\text{\tiny{NL}})}}{4} + \frac{\mathcal{J}}{8} \, x_{(\text{\tiny{L}})} \cdot& x_{(\text{\tiny{NL}})} + \frac{9}{64} x_{(\text{\tiny{L}})}^2 + \frac{\mathcal{J}}{16} x_{(\text{\tiny{L}})}^3 \Bigg\} = - \frac{2\sqrt{\lambda}}{\pi} \frac{1}{16\mathcal{J}^2}\left(W^2 + W^3\right) \Rightarrow \nonumber \\[12pt]
\Rightarrow &E - J\Big|_{(\text{NL})} &= - \frac{\sqrt{\lambda}}{32\,\pi} \sum_{n = 1}^\infty \Bigg\{16 \, \beta_n + \sum_{k = 1}^{n-1} &\alpha_k \bigg[9 \, \alpha_{n - k} + 8 \, \beta_{n - k}\bigg] + \nonumber \\[6pt]
&&& + 4 \sum_{k,m = 1}^{n - 2} \alpha_{k} \, \alpha_{m} \, \alpha_{n - k - m}\Bigg\} \cdot \mathcal{J}^{n-2} \left(e^{-2\mathcal{J} - 2}\right)^{n}. \label{AnomalousDimensions12} \qquad
\end{IEEEeqnarray} \normalsize
\subsection[NNL Terms]{NNL Terms \label{NexttosubleadingII}}
Computing higher-order terms in the long-string expansion of $E - J$ is straightforward. Equation \eqref{xEquationII2} becomes,
\footnotesize \begin{IEEEeqnarray}{c}
x_{(\text{\tiny{L}} + \text{\tiny{NL}} + \text{\tiny{NNL}} + \ldots)} = \sum_{n = 1}^{\infty} \frac{x_0^n}{n!} \cdot \sum_{k,j = 0}^{n - 1} \frac{\left(-1\right)^k n^k \left(n - 1\right)!}{\left(k - j_1 - j_2\right)! \, j_1! \, j_2!} \left(\frac{\mathcal{J} + 1}{2}\right)^{k - j_1 - j_2} \left(\frac{14\mathcal{J} + 9}{64}\right)^{j_1} \left(\frac{15\mathcal{J}}{128}\right)^{j_2}, \qquad
\end{IEEEeqnarray} \normalsize
with $n - 1 = k + j_1 + 2j_2$ and $0 \leq j_1 + j_2 \leq k$. Again we have to select and keep only the leading, subleading (NL) and next-to-subleading (NNL) terms the way it was done in the previous subsection and then express the resulting power series with the aid of Lambert's function, by using the formulas of appendix \ref{LambertAppendix}. We find:
\begin{IEEEeqnarray}{ll}
x_{(\text{\tiny{L}} + \text{\tiny{NL}} + \text{\tiny{NNL}})} & = \sum_{n = 1}^{\infty} \left(\alpha_n \, \mathcal{J}^{n - 1} + \beta_n \, \mathcal{J}^{n - 2} + \gamma_n \, \mathcal{J}^{n - 3}\right) \cdot \left(e^{-2\mathcal{J} - 2}\right)^{n}, \label{Leading+Subleading+SubSubleading_x1}
\end{IEEEeqnarray}
having defined ($\alpha$'s and $\beta$'s are defined in \eqref{AlphaCoefficients}-\eqref{BetaCoefficients}) the $\gamma_n$'s as
\begin{IEEEeqnarray}{l}
\gamma_n \equiv \left(- 1\right)^{n + 1} \, 2^{3n - 6} \cdot \frac{n^{n - 3}}{n!} \cdot \left(n - 1\right)\left(n - 2\right) \left(n^2 + 41n + 228\right). \label{GammaCoefficients}
\end{IEEEeqnarray} \normalsize
\indent In terms of Lambert's W-function, the inverse spin function $x = x\left(\mathcal{J}\right)$ (precise up to NNL order) is given by
\small\begin{IEEEeqnarray}{ll}
x_{(\text{\tiny{L}} + \text{\tiny{NL}} + \text{\tiny{NNL}})} = \frac{2}{\mathcal{J}} \, W - \frac{1}{4\,\mathcal{J}^2}\frac{W^2 \left(7\,W + 8\right)}{1 + W} + \frac{1}{64\,\mathcal{J}^3} \frac{W^3 \left(76 W^3 + 269 W^2 + 312 W + 120\right)}{\left(1 + W\right)^3}, \qquad \label{Leading+Subleading+Nexttosubleading_x}
\end{IEEEeqnarray} \normalsize
where the arguments of the W-functions are $W\left(8\,\mathcal{J}\,e^{- 2\mathcal{J} - 2}\right)$. This expression for $x$ is in turn inserted into \eqref{AnomalousDimensions4}, keeping only terms up to next-to-subleading order. The next-to-subleading (NNL) coefficients of $E - J$ are found by writing
\small\begin{IEEEeqnarray}{ll}
E - J\Big|_{(\text{NNL})} &= - \frac{2\sqrt{\lambda}}{\pi} \Bigg\{\frac{x_{(\text{\tiny{NNL}})}}{4} + \frac{9}{32} \, x_{(\text{\tiny{L}})} \cdot x_{(\text{\tiny{NL}})} + \frac{\mathcal{J}}{16} \, x_{(\text{\tiny{NL}})}^2 + \frac{\mathcal{J}}{8} \, x_{(\text{\tiny{L}})} \cdot x_{(\text{\tiny{NNL}})} + \frac{23}{256} \, x_{(\text{\tiny{L}})}^3 + \nonumber \\[6pt]
& + \frac{3\,\mathcal{J}}{16} \, x_{(\text{\tiny{L}})}^2 \cdot x_{(\text{\tiny{NL}})} + \frac{111\,\mathcal{J}}{2048} \, x_{(\text{\tiny{L}})}^4\Bigg\} = - \frac{2\sqrt{\lambda}}{\pi} \frac{1}{256\mathcal{J}^3}\frac{W^3\left(11\,W^2 + 26\,W + 16\right)}{1 + W}, \qquad \label{AnomalousDimensions14}
\end{IEEEeqnarray} \normalsize
which implies,
\small\begin{IEEEeqnarray}{ll}
E - J&\Big|_{(\text{NNL})} = - \frac{\sqrt{\lambda}}{128\,\pi} \sum_{n = 1}^\infty \Bigg\{64 \, \gamma_n + 8 \sum_{k = 1}^{n-1} \bigg[9 \, \alpha_k \, \beta_{n - k} + 2 \, \beta_k \, \beta_{n - k} + 4 \, \alpha_k \, \gamma_{n - k}\bigg] + \sum_{k,m = 1}^{n - 2} \alpha_{k} \, \alpha_{m} \cdot \nonumber \\[6pt]
& \cdot \bigg[23 \, \alpha_{n - k - m} + 48 \, \beta_{n - k - m}\bigg] + \frac{111}{8} \sum_{k,m,s = 1}^{n - 3} \alpha_{k} \, \alpha_{m} \alpha_{s} \, \alpha_{n - k - m - s}\Bigg\} \cdot \mathcal{J}^{n-3} \left(e^{-2\mathcal{J} - 2}\right)^{n}. \qquad \label{AnomalousDimensions15}
\end{IEEEeqnarray} \normalsize
\indent Our final result for the inverse spin function and the anomalous dimensions of the long $\mathcal{N} = 4$ SYM operators $\text{Tr}\left[\Phi\mathcal{Z}^m \, \Phi \, \mathcal{Z}^{J-m}\right] + \ldots$ at strong 't Hooft coupling and up to next-to-subleading order in large-$\mathcal{J}$ is:
\small\begin{IEEEeqnarray}{l}
x = \frac{2\,W}{\mathcal{J}} - \frac{1}{4\,\mathcal{J}^2}\frac{W^2 \left(7\,W + 8\right)}{1 + W} + \frac{1}{64\,\mathcal{J}^3} \frac{W^3 \left(76 W^3 + 269 W^2 + 312 W + 120\right)}{\left(1 + W\right)^3} + \ldots \qquad \label{InverseSpinFunctionII1}\\[6pt]
\mathcal{E} - \mathcal{J} = 1 - \frac{1}{4\,\mathcal{J}}\left(2 W + W^2\right) - \frac{1}{16\,\mathcal{J}^2}\left(W^2 + W^3\right) - \frac{1}{256\,\mathcal{J}^3}\frac{W^3\left(11\,W^2 + 26\,W + 16\right)}{1 + W} + \ldots, \qquad \; \label{AnomalousDimensions16}
\end{IEEEeqnarray} \normalsize
where $\mathcal{E} \equiv \pi E / 2\sqrt{\lambda}$, $\mathcal{J} \equiv \pi J / 2\sqrt{\lambda}$ and the argument of the W-function is $W\left(8\,\mathcal{J}\,e^{- 2\mathcal{J} - 2}\right)$. Expanding series \eqref{InverseSpinFunctionII1} and \eqref{AnomalousDimensions16} around $\mathcal{J} \rightarrow \infty$, by using the formulas of appendix \ref{LambertAppendix}, we find that they completely agree with the ones computed if series \eqref{GKPIISpin2} is inverted with a symbolic calculations program such as Mathematica and the result is plugged into equation \eqref{GKPIIEnergy2} (cf. appendix \ref{Long-Fast-StringsAppendix}, equations \eqref{Exact_Value_of_xII1}--\eqref{MathematicaAnomalousDimensionsII1}). \\
\indent Likewise, we may keep on going to higher and higher orders in $J$. Theoretically, we are thus able to obtain all the terms of the long string expansion. We may conjecture that the Lambert W-functions will keep appearing to all orders of $x$ and consequently to all orders of $\mathcal{E} - \mathcal{J}$ as well. To see this, note that equation \eqref{xEquationII2} will in general contain a term of the form $n^n/n!$ that multiplies some Laurent polynomial of $n$, which in turn originates from the multinomial coefficient and the expansion of a$_i$'s in powers of $\mathcal{J}$. Using the formulas of appendix \ref{LambertAppendix}, the resulting series may be expressed in terms of W-functions. \\
\subsection[Fast Circular Strings]{Fast Circular Strings in $\mathbb{R}\times\text{S}^2$: $\omega \rightarrow 1^-$, $J \gg \lambda$ \label{FastCircularStrings}}
Region $\omega < 1$ of GKP strings on the sphere (circular strings) is very similar to the regime $\omega >1$ that was studied in sections \ref{GKPStringII} and \ref{InverseSpinFunctionII}-\ref{NexttosubleadingII}. In this subsection we will briefly derive the corresponding expressions $\widetilde{x} = \widetilde{x}\left(\mathcal{J}\right)$ and $\mathcal{E} = \mathcal{E}\left(\mathcal{J}\right)$ for fast (large-J) circular strings on S$^2$ ($\omega \rightarrow 1^{-}$). Our treatment is very similar to the case $\omega \rightarrow 1^+$:
\small\begin{IEEEeqnarray}{c}
\mathcal{E} \equiv \frac{\pi\,E}{2\sqrt{\lambda}} = \sum_{n = 0}^{\infty} \widetilde{x}^n\left(d_n\ln \widetilde{x} + h_n\right) \qquad \label{GKPIICircularEnergy2} \\[6pt]
\mathcal{J} \equiv \frac{\pi\,J}{2\sqrt{\lambda}} = \frac{1}{\sqrt{1 - \widetilde{x}}} \cdot \sum_{n = 0}^{\infty} \widetilde{x}^n\left(c_n\ln \widetilde{x} + b_n\right) = \sum_{n = 0}^{\infty} \widetilde{x}^n \cdot \sum_{k = 0}^n \frac{(2k - 1)!!}{\left(2k\right)!!} \left(c_{n - k} \ln \widetilde{x} + b_{n - k}\right) \label{GKPIICircularSpin2} \\[6pt]
\mathcal{E} - \mathcal{J} = \sum_{n = 0}^{\infty} \widetilde{x}^n \left(f_n \, \ln \widetilde{x} + g_n\right) = \sum_{n = 0}^{\infty} \widetilde{x}^n \left[A_n + f_n \, \ln \frac{\widetilde{x}}{x_0}\right], \qquad \label{GKPIICircularAnomalousDimensions1}
\end{IEEEeqnarray} \normalsize
where the complementary parameter is now $\widetilde{x} \equiv 1 - \omega^2 \rightarrow 0^{-}$, $b_n$, $c_n$, $d_n$, $h_n$ are defined in \eqref{LongSeriesCoefficientsII1},
\begin{IEEEeqnarray}{c}
f_n \equiv d_n - \sum_{k = 0}^n \frac{(2k - 1)!!}{\left(2k\right)!!} \cdot c_{n - k}\,, \quad g_n \equiv h_n - \sum_{k = 0}^n \frac{(2k - 1)!!}{\left(2k\right)!!} \cdot b_{n - k}
\end{IEEEeqnarray}
and
\begin{IEEEeqnarray}{c}
A_n \equiv g_n + f_n \ln x_0 = g_n + 2 f_n \, \left(2\ln 2 - \mathcal{J} - 1\right), \qquad
\end{IEEEeqnarray}
while $x_0$ is defined in \eqref{DefinitionIIx_0} and $n = 0 \,, 1 \,, 2 \,, \ldots$ We find: \\
\small\begin{IEEEeqnarray}{l}
\widetilde{x} = -\frac{2\,W}{\mathcal{J}} - \frac{1}{4\,\mathcal{J}^2}\frac{W^2 \left(9\,W + 8\right)}{1 + W} - \frac{1}{64\,\mathcal{J}^3} \frac{W^3 \left(140 W^3 + 397 W^2 + 376 W + 120\right)}{\left(1 + W\right)^3} + \ldots \qquad \label{CircularInverseSpinFunctionII1} \\[6pt]
\mathcal{E} - \mathcal{J} = 1 - \frac{1}{4\,\mathcal{J}}\left(2 W + W^2\right) - \frac{1}{16\,\mathcal{J}^2}\left(W^2 + W^3\right) - \frac{1}{256\,\mathcal{J}^3}\frac{W^3\left(11\,W^2 + 26\,W + 16\right)}{1 + W} + \ldots, \qquad \label{GKPIICircularAnomalousDimensions2}
\end{IEEEeqnarray} \normalsize \\
where the argument of the W-function is $W\left(- 8 \, \mathcal{J} \, e^{-2\mathcal{J} - 2}\right)$. Notice that although the inverse spin functions $x\left(\mathcal{J}\right)$ and $\widetilde{x}\left(\mathcal{J}\right)$ are different for fast folded and circular strings on S$^2$ (cf. \eqref{InverseSpinFunctionII1}, \eqref{CircularInverseSpinFunctionII1}), the expressions for the anomalous dimensions in terms of Lambert W-functions coincide (cf. \eqref{AnomalousDimensions16}, \eqref{GKPIICircularAnomalousDimensions2}). Nevertheless, and due to the fact that the arguments of the W-functions have opposite signs in the two cases, the formulas for the anomalous dimensions $\gamma = \gamma\left(\mathcal{J}\right)$ will have a periodic sign difference (cf. \eqref{MathematicaAnomalousDimensionsII1}-\eqref{MathematicaAnomalousDimensionsII2}). It would be interesting to study the operators that are dual to these circular string states. \\
\section[Gubser-Klebanov-Polyakov AdS$_3$ String]{Gubser-Klebanov-Polyakov AdS$_3$ String \label{GKPStringI}}
\begin{figure}
\begin{center}
\includegraphics[scale=0.4]{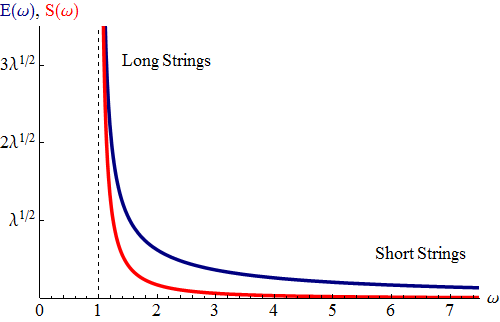}
\qquad
\includegraphics[scale=0.4]{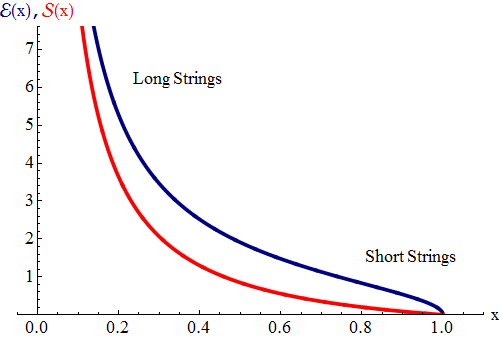}
\caption{Energy and spin of the folded closed AdS$_3$ string as functions of $\omega > 1$ and $x > 0$.} \label{Graph:GKPIb}
\end{center}
\end{figure}

\begin{figure}
\begin{center}
\includegraphics[scale=0.4]{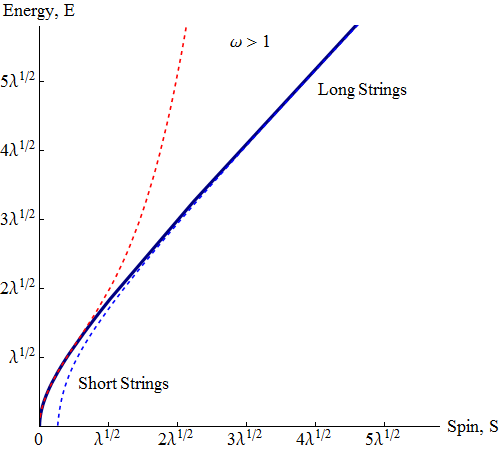}
\caption{Energy vs spin of the folded closed AdS$_3$ string for $\omega > 1$.} \label{Graph:Energy-SpinIb}
\end{center}
\end{figure}

Let us now consider the Gubser-Klebanov-Polyakov (GKP) folded closed string that rotates at the equator of $\text{S}^3$ of AdS$_5$ \cite{GubserKlebanovPolyakov02}:
\begin{IEEEeqnarray}{c}
\Big\{t = \kappa \tau, \, \rho = \rho(\sigma), \, \theta = \kappa \omega \tau, \, \phi_1 = \phi_2 = 0\Big\} \times \Big\{\overline{\theta}_1 = \overline{\theta}_2 = \overline{\phi}_1 = \overline{\phi}_2 = \overline{\phi}_3 = 0\Big\}. \qquad \label{GKPAnsatzI1}
\end{IEEEeqnarray}
\indent As in the case of the GKP string rotating in $\mathbb{R}\times\text{S}^2$, we will show that the GKP solution in AdS$_3$ is unique in the sense that it minimizes the energy of an AdS$_3$ string with a single spin $S$. Let us again consider the following generic ansatz:
\begin{IEEEeqnarray}{c}
\Big\{t = \kappa \tau, \, \rho = \rho(\tau,\sigma), \, \theta = \theta\left(\tau,\sigma\right), \, \phi_1 = \phi_2 = 0\Big\} \times \Big\{\overline{\theta}_1 = \overline{\theta}_2 = \overline{\phi}_1 = \overline{\phi}_2 = \overline{\phi}_3 = 0\Big\}. \qquad \label{GKPAnsatzI2}
\end{IEEEeqnarray}
The Polyakov action in the conformal gauge ($\gamma_{ab} = \eta_{ab}$) is:
\begin{IEEEeqnarray}{c}
\mathcal{S}_P = \frac{\ell^2}{4\pi \alpha'}\int \left(- \dot{t}^2 \cosh^2\rho + \dot{\rho}^2 - \rho'^2 + \big(\dot{\theta}^2 - \theta'^2\big)\sinh^2\rho\right) d\tau d\sigma.
\end{IEEEeqnarray}
From this action one can derive the Hamiltonian density of the string that satisfies the first of the Virasoro constraints, namely
\begin{IEEEeqnarray}{c}
\mathcal{H} = \frac{\ell^2}{4\pi\alpha'} \left\{- \kappa^2 \cosh^2\rho + \frac{p_{\rho}^2}{4} + \rho'^2 + \frac{p_{\theta}^2}{4\sinh^2\rho} + \theta'^2\,\sinh^2\rho\right\} = 0\,, \quad \begin{array}{l} p_{\rho} = 2\,\dot{\rho} \\ p_{\theta} =
2\,\dot{\theta}\,\sinh^2\rho. \end{array} \qquad \label{VirasoroI1}
\end{IEEEeqnarray}
\noindent We want to find the minimum of the energy $E$ subject to the constraint of fixed angular momentum $S$. Introducing the Lagrange multiplier $\omega$, the functional to be minimized is given by:
\begin{IEEEeqnarray}{c}
Y \equiv \frac{\ell^2}{4\pi \alpha'}\int d\sigma \cdot 2\cosh\rho\,\sqrt{\frac{p_{\rho}^2}{4} + \rho'^2 +
\frac{p_{\theta}^2}{4\sinh^2\rho} + \theta'^2 \, \sinh^2\rho} - \omega \left[\frac{\ell^2}{4\pi\alpha'} \int d\sigma \, p_{\theta} -S\right], \qquad
\end{IEEEeqnarray}
where the last formula is obtained by substituting $\kappa$ from \eqref{VirasoroI1} into the integral of energy $E$. Therefore,
\begin{IEEEeqnarray}{c}
\frac{\delta Y}{\delta p_{\rho}} = 0 \Rightarrow p_{\rho} = 0\Rightarrow \dot{\rho} = 0 \Rightarrow \rho = \rho(\sigma)\label{GKPMinimumI1} \\[6pt]
\frac{\delta Y}{\delta p_{\theta}} = 0 \Rightarrow \dot{\theta} = \kappa\,\omega \Rightarrow \theta = \kappa\,\omega\,\tau + \widetilde{\theta}(\sigma).
\end{IEEEeqnarray}
\indent Combining the second Virasoro constraint $\dot{\rho} \, \rho' + \dot{\theta}\, \theta' \, \sinh^2\rho = 0$ with equation \eqref{GKPMinimumI1}, we conclude that $\theta'=0$, i.e.\ $\widetilde{\theta}(\sigma)=0$. Again, the vanishing of the functional derivative $\delta Y/\delta\rho(\sigma) = 0$ will give the equation of motion for $\rho$. Thus the GKP solution \eqref{GKPAnsatzI1} follows from the minimization of the string energy in $\text{AdS}_3$. Its Polyakov action reads:
\begin{IEEEeqnarray}{ll}
\mathcal{S}_P &= \frac{\ell^2}{4 \pi \alpha'} \int \left(-\dot{t}^2 \cosh^2\rho - \rho'\,^2 + \dot{\theta}^2 \sinh^2\rho\right) d\tau d\sigma = \nonumber \\[6pt]
& = \frac{\ell^2}{4 \pi \alpha'} \int \left(-\kappa^2 \cosh^2\rho - \rho'\,^2 + \kappa^2 \omega^2 \sinh^2\rho\right) d\tau d\sigma, \label{GKPActionI}
\end{IEEEeqnarray}
\begin{figure}
\begin{center}
\includegraphics[scale=0.4]{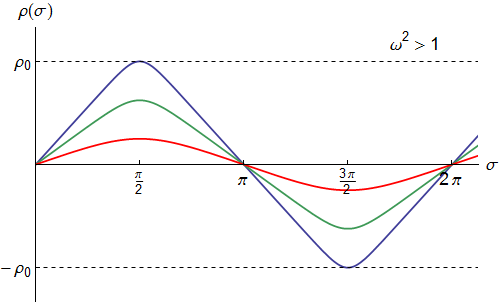}
\qquad
\includegraphics[scale=0.4]{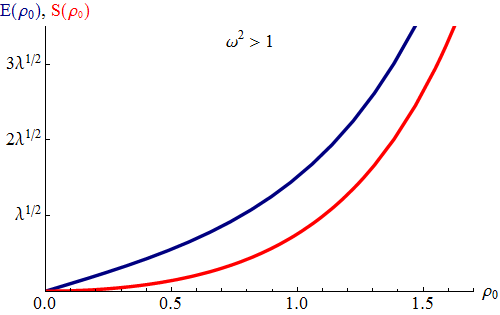}
\caption{\small{$\rho = \rho\left(\sigma\right)$ and energy/spin of the folded closed string AdS$_3$ string \eqref{GKPAnsatzI1} for $\omega > 1$.}} \label{Graph:GKPIa}
\end{center}
\end{figure}
where $\ell^4 / \alpha'\,^2 = \lambda$ is the 't Hooft coupling. The string essentially contains four segments extending between $\rho = 0$ and $\rho = \rho_0$. $\kappa$ is a factor needed to fix $\sigma\left(\rho_0\right) = \pi/2$. The conserved charges that correspond to the two cyclic coordinates t and $\theta$ of the action \eqref{GKPActionI} both diverge when $\omega<1$. For $\omega >1$ we obtain: \\[6pt]
\noindent$\underline{\omega^2 > 1}$. Folded closed string.
\begin{IEEEeqnarray}{l}
\rho(\sigma) = \text{arctanh} \left[\frac{1}{\omega} sn\left(\kappa \omega \sigma \, \Bigg| \,\frac{1}{\omega^2}\right)\right], \quad \kappa = \frac{2}{\pi\omega} \cdot \mathbb{K}\left(\frac{1}{\omega^2}\right), \quad \omega = \coth\rho_0 \qquad \label{GKPILength2}
\end{IEEEeqnarray}
\begin{IEEEeqnarray}{l}
E(\omega) = \frac{2 \sqrt{\lambda}}{\pi} \, \frac{\omega}{\omega^2 - 1} \cdot \mathbb{E} \left(\frac{1}{\omega^2}\right) \Rightarrow \mathcal{E} \equiv \frac{\pi \, E}{2 \, \sqrt{\lambda}} = \frac{\sqrt{1 - x}}{x} \cdot \mathbb{E}\left(1 - x\right) \label{GKPIEnergy2}
\end{IEEEeqnarray}
\begin{IEEEeqnarray}{l}
S(\omega) = \frac{2 \sqrt{\lambda}}{\pi} \, \left[\frac{\omega^2}{\omega^2 - 1} \mathbb{E} \left(\frac{1}{\omega^2}\right) - \mathbb{K} \left(\frac{1}{\omega^2}\right)\right] \Rightarrow \mathcal{S} \equiv \frac{\pi \, S}{2 \, \sqrt{\lambda}} = \frac{1}{x} \, \mathbb{E}\left(1 - x\right) - \mathbb{K}\left(1 - x\right) \qquad \label{GKPISpin2}
\end{IEEEeqnarray}
\begin{IEEEeqnarray}{l}
\gamma \equiv \mathcal{E} - \mathcal{S} = \frac{\sqrt{1 - x} - 1}{x} \cdot \mathbb{E}\left(1 - x\right) + \mathbb{K}\left(1 - x\right),
\end{IEEEeqnarray}
where $x \equiv 1 - 1/\omega^2$ is the complementary parameter of $1/\omega^2$. The plot of $\rho\left(\sigma\right)$ for various values of the angular velocity $\omega > 1$ as well as the string's energy and spin as functions of $\rho_0$, $\omega$ and $x$ may be found in figures \ref{Graph:GKPIa}--\ref{Graph:GKPIb}. In figure \ref{Graph:Energy-SpinIb} we have plotted the energy of the string as a function of its spin $E = E(S)$ parametrically, along with the first 4 terms of its "short" series (red dashed line, equation \eqref{ShortStringIEnergy3}) and its leading "long" approximation (blue dashed line, equations \eqref{GKPILeading1}--\eqref{GKPILeading2}).
\\
\subsection[Long Strings]{Long Strings: $\omega \rightarrow 1^+$, $S \gg \sqrt{\lambda}$}
In the long-strings regime, $\omega \rightarrow 1^+$ ($S \gg \lambda$), the formulas for the energy and spin have the following expansions:
\small \begin{IEEEeqnarray}{lll}
E & = \frac{2 \sqrt{\lambda}}{\pi\omega} \cdot \bigg\{&\frac{\omega^2}{\omega^2 - 1} + \frac{1}{2 \pi} \sum_{n = 0}^\infty \frac{\Gamma(n + 1/2) \Gamma(n + 3/2)}{n!(n + 1)!} \, \Big[2\psi(n + 1) - 2\psi(n + 1/2) - \nonumber \\[6pt]
&& - \ln(1 - 1/\omega^2) - \frac{1}{\left(n + 1\right)\left(2n + 1\right)}\Big] \cdot (1 - 1/\omega^2)^n \bigg\} \qquad \label{GKPIEnergy3} \\[12pt]
S & = \frac{2 \sqrt{\lambda}}{\pi} \cdot \bigg\{&\frac{\omega^2}{\omega^2 - 1} - \frac{1}{4 \pi} \sum_{n = 0}^\infty \frac{\left(\Gamma(n + 1/2)\right)^2}{n!(n + 1)!} \, \Big[2\psi(n + 1) - 2\psi(n + 1/2) - \ln(1 - 1/\omega^2) + \nonumber \\[6pt]
&& + \frac{1}{n + 1}\Big] \cdot (1 - 1/\omega^2)^n \bigg\}. \label{GKPISpin3}
\end{IEEEeqnarray} \normalsize
The two series may also be written in terms of the complementary parameter $x \equiv 1 - 1 / \omega^2 \rightarrow 0^+$:
\begin{IEEEeqnarray}{ll}
\mathcal{E} \equiv \frac{\pi\,E}{2\sqrt{\lambda}} &= \sqrt{1 - x} \cdot \left\{\frac{1}{x} + \sum_{n = 0}^{\infty} x^n\left(d_n\ln x + h_n\right)\right\} = \nonumber \\[6pt]
& = \frac{1}{x} - \sum_{n = 0}^{\infty} x^n \cdot \left\{\frac{\left(2n - 1\right)!!}{\left(2n + 2\right)!!} + \sum_{k = 0}^n \frac{(2k - 3)!!}{\left(2k\right)!!} \left(d_{n - k} \ln x + h_{n - k}\right)\right\} \qquad \label{GKPIEnergy4} \\[12pt]
\mathcal{S} \equiv \frac{\pi\,S}{2\sqrt{\lambda}} &= \frac{1}{x} + \sum_{n = 0}^{\infty} x^n\left(c_n\ln x + b_n\right). \label{GKPISpin4}
\end{IEEEeqnarray}
The coefficients that appear in series \eqref{GKPIEnergy4} and \eqref{GKPISpin4} are given by:\footnote{As a reminder, $0!! = 1$, $\left(-1\right)!! = 1$, $\left(-3\right)!! = -1$.}
\begin{IEEEeqnarray}{c}
d_n = - \frac{1}{4} \left(\frac{\left(2n - 1\right)!!}{\left(2n\right)!!}\right)^2 \cdot \frac{2n + 1}{n + 1} \,, \qquad h_n = - d_n \cdot \left[4\ln2 + 4\left(H_n - H_{2n}\right) + \frac{1}{n + 1} - \frac{2}{2n + 1}\right] \nonumber \\[12pt]
c_n = - \frac{d_n}{2n + 1} \,, \qquad b_n = - c_n \cdot \left[4\ln2 + 4\left(H_n - H_{2n}\right) + \frac{1}{n + 1}\right], \label{LongSeriesCoefficientsI1}
\end{IEEEeqnarray}
where $n = 0 \,, 1 \,, 2 \,, \ldots$ \\
\subsection[Short-Long Strings Duality]{Short-Long Strings Duality \label{ShortLongDualityI}}
Solving \eqref{GKPIEnergy2} and \eqref{GKPISpin2} for $\mathbb{E}(k)$ and $\mathbb{K}(k)$ and substituting into Legendre's relation \eqref{Legendre}, we get the following formula between classical folded short and long strings spinning in AdS$_3 \subset \text{AdS}_5$ \cite{GeorgiouSavvidy11}:
\begin{IEEEeqnarray}{c}
\frac{1}{\omega} E S' + \frac{1}{\omega'} E' S - S S' = \frac{2 \lambda}{\pi}, \label{Short-Long2a}
\end{IEEEeqnarray}
where the respective arguments of the primed and unprimed charges are $k = 1 / \omega^2$ and $k' = x = 1 / \omega' \, ^2$, and satisfy $k + k' = 1$. Again we see that large values of $\omega' \rightarrow \infty$ ("short" strings) correspond to values of $\omega \rightarrow 1^+$ near unity ("long" strings) and \eqref{Short-Long2a} provides a map between the corresponding energies and spins.\footnote{This subsection has been included here for reasons of completeness. For more, see \cite{GeorgiouSavvidy11}. One may also formulate short-long dualities that mix the charges of the AdS$_3$ and $\mathbb{R}\times\text{S}^2$ strings:
\begin{IEEEeqnarray}{c}
\frac{\omega'}{\omega}\,\mathcal{E}'_1\,\mathcal{E}_2 + \frac{\omega}{\omega'}\,\mathcal{E}_1\,\mathcal{E}'_2 - \omega \, \omega' \,\mathcal{E}_2\,\mathcal{E}'_2 = \frac{\pi}{2} \quad \& \quad \mathcal{S}_1\,\mathcal{J}'_2 + \mathcal{S}'_1\,\mathcal{J}_2 + \mathcal{S}_1\,\mathcal{S}'_1 = \frac{\pi}{2}\\
\frac{1}{\omega'}\,\mathcal{S}_1\,\mathcal{E}'_2 + \frac{1}{\omega}\,\mathcal{S}'_1\,\mathcal{E}_2 = \frac{\pi}{2} \quad \& \quad \frac{1}{\omega\,\omega'}\,\mathcal{E}_1\,\mathcal{E}'_1 - \mathcal{J}_2\,\mathcal{J}'_2 = \frac{\pi}{2},
\end{IEEEeqnarray}
where the index 1 refers to the string energy and spin in AdS$_3$ and index 2 denotes the corresponding charges in $\mathbb{R}\times\text{S}^2$.} \\
\section[Inverse Spin Functions and Anomalous Dimensions on $\text{AdS}_3$]{Inverse Spin Functions and Anomalous Dimensions on $\text{AdS}_3$ \label{SpinFunctionI}}
\subsection[Inverse Spin Function]{Inverse Spin Function \label{InverseSpinFunctionI}}
Following the method of subsection \ref{InverseSpinFunctionII}, we will now invert series \eqref{GKPISpin4} for $x = x\left(\mathcal{S}\right)$. First solve \eqref{GKPISpin4} for $\ln x$:
\begin{IEEEeqnarray}{ll}
\mathcal{S} = \frac{1}{x} + \sum_{n = 0}^{\infty} &x^n\left(c_n\ln x + b_n\right) \Rightarrow \nonumber \\
& \Rightarrow \ln x = \left[-\frac{1}{c_0 \, x} + \frac{\mathcal{S} - b_0}{c_0} - \sum_{n = 1}^{\infty} \frac{b_n}{c_0} \, x^n\right] \cdot \sum_{n = 0}^{\infty} \left(-1\right)^n \left(\sum_{k = 1}^{\infty} \frac{c_k}{c_0} \, x^k\right)^{n}. \label{InverseSpinEquationI}\qquad
\end{IEEEeqnarray} \\
This essentially reduces to an equation of the following form (cf. equation \eqref{xEquationII1}):
\begin{IEEEeqnarray}{c}
x = x_0 \cdot \exp\left[\frac{\text{a}_0}{x} + \sum_{n = 1}^{\infty} \text{a}_n \, x^n\right] = x_0 \cdot \exp\left(\frac{\text{a}_0}{x} + \text{a}_1 \, x + \text{a}_2 \, x^2 + \text{a}_3 \, x^3 + \ldots\right), \qquad \label{xEquationI1}
\end{IEEEeqnarray}
where the a$_n$'s are linear functions of $\mathcal{S}$ determined from \eqref{InverseSpinEquationI} (a$_0 = -c_0^{-1} = -4$), and $x_0$ is defined as:
\begin{IEEEeqnarray}{c}
x_0 \equiv \exp\left[\frac{\mathcal{S} - b_0}{c_0} + \frac{c_1}{c_0^2}\right] = 16 \, e^{4\mathcal{S} + 3/2}.
\end{IEEEeqnarray}
\indent Equation \eqref{xEquationI1}, that gives the inverse spin function of a string that rotates in AdS$_3$ differs significantly from the corresponding inverse spin function in S$^5$, \eqref{xEquationII1}. For one, it contains the $1/x$ term but also $x_0$ is now an increasing function of $\mathcal{S}$. The point is that we cannot solve equation \eqref{xEquationI1} by the algorithm of subsection \ref{InverseSpinFunctionII}. Instead, a slightly different procedure has to be applied. Suppose $x^*$ solves the following equation:
\begin{IEEEeqnarray}{c}
x^* = x_0 \cdot e^{\text{a}_0 / x^*} \Rightarrow x^* = \frac{\text{a}_0}{W\left(\text{a}_0 / x_0\right)} = x_0 \cdot e^{W\left(\text{a}_0 / x_0\right)}, \label{xstar1}
\end{IEEEeqnarray}
where $W\left(z\right)$ is again the Lambert W-function (see \eqref{LambertDefinition2} in appendix \ref{LambertAppendix}). Essentially $x^*$ is the leading-order solution of equation \eqref{xEquationI1}.\footnote{Observe though that since we solve \eqref{InverseSpinEquationI} in the region where $\mathcal{S} \rightarrow +\infty$, $x\rightarrow0^+$ and a$_0 < 0$, $x_0 \rightarrow +\infty$, we must choose the $W_{-1}$ branch of Lambert's function. We'll return to this issue later.} Now set:
\begin{IEEEeqnarray}{c}
x = x^* \cdot e^v, \label{xstar2}
\end{IEEEeqnarray}
\noindent with $v\rightarrow 0$, and substitute it into equation \eqref{xEquationI1}. We obtain, using also \eqref{xstar1}:
\begin{IEEEeqnarray}{c}
v - \frac{\text{a}_0}{x^*} \, \sum_{k = 1}^{\infty} \left(-1\right)^k \frac{v^k}{k!} - \sum_{n = 1}^{\infty} \text{a}_n \, \left(x^*\right)^n \, e^{nv} = 0. \label{xstar3}
\end{IEEEeqnarray}
\indent This series may be inverted for \textit{v} using the standard series inversion. First we expand the exponential in \eqref{xstar3}:
\small\begin{IEEEeqnarray}{c}
\left(1 + \frac{\text{a}_0}{x^*} - \sum_{k = 1}^{\infty} k \, \text{a}_k \, \left(x^*\right)^k\right) v - \sum_{n = 2}^{\infty} \left[\left(-1\right)^n \frac{\text{a}_0}{x^*} + \sum_{k = 1}^{\infty} k^n \, \text{a}_k \, \left(x^*\right)^k\right] \frac{v^n}{n!} = \sum_{n = 1}^{\infty} \text{a}_n \, \left(x^*\right)^n. \qquad
\end{IEEEeqnarray} \normalsize
The inverse series is a power series in $x^*$. It can be obtained by using a symbolic calculations package such as Mathematica. The result is:
\begin{IEEEeqnarray}{c}
v = \frac{\text{a}_1}{\text{a}_0}\left(x^*\right)^2 + \left[\frac{\text{a}_2}{\text{a}_0} - \frac{\text{a}_1}{\text{a}_0^2}\right]\left(x^*\right)^3 + \left[\frac{\text{a}_1}{\text{a}_0^3} + \frac{3 \, \text{a}_1^2 - 2 \, \text{a}_2}{2 \, \text{a}_0^2} + \frac{\text{a}_3}{\text{a}_0}\right] \left(x^*\right)^4 + \ldots \qquad
\end{IEEEeqnarray}
Then $x$ in \eqref{xstar2} is given by
\begin{IEEEeqnarray}{c}
x = x^* + \frac{\text{a}_1}{\text{a}_0}\left(x^*\right)^3 + \left[\frac{\text{a}_2}{\text{a}_0} - \frac{\text{a}_1}{\text{a}_0^2}\right]\left(x^*\right)^4 + \left[\frac{\text{a}_3}{\text{a}_0} + \frac{2 \, \text{a}_1^2 - \text{a}_2}{\text{a}_0^2} + \frac{\text{a}_1}{\text{a}_0^3}\right] \left(x^*\right)^5 + \ldots \qquad \label{xI1}
\end{IEEEeqnarray}
and $1 / x$ by
\begin{IEEEeqnarray}{c}
\frac{1}{x} = \frac{1}{x^*} - \frac{\text{a}_1}{\text{a}_0} x^* - \left[\frac{\text{a}_2}{\text{a}_0} - \frac{\text{a}_1}{\text{a}_0^2}\right] \left(x^*\right)^2 - \left[\frac{\text{a}_3}{\text{a}_0} + \frac{\text{a}_1^2 - \text{a}_2}{\text{a}_0^2} + \frac{\text{a}_1}{\text{a}_0^3}\right] \left(x^*\right)^3 + \ldots \qquad \label{xI2}
\end{IEEEeqnarray}
\indent We may now expand \eqref{InverseSpinEquationI}, obtain the coefficients a$_n$ and, by substituting them into \eqref{xI1}--\eqref{xI2}, find the corresponding expressions for $x$ and $1/x$ (only the first few terms are shown):
\begin{IEEEeqnarray}{l}
x = x^* + \left(\frac{S}{8} + \frac{3}{64}\right) \left(x^*\right)^3 + \left(\frac{S}{16} + \frac{23}{1024}\right) \left(x^*\right)^4 + \left(\frac{S^2}{32} + \frac{55 S}{1024} + \frac{349}{24\,576}\right) \left(x^*\right)^5 + \ldots \qquad \label{xI3} \\[12pt]
\frac{1}{x} = \frac{1}{x^*} - \left(\frac{S}{8} + \frac{3}{64}\right) x^* - \left(\frac{S}{16} + \frac{23}{1024}\right) \left(x^*\right)^2 - \left( \frac{S^2}{64} + \frac{43 S}{1024} + \frac{295}{24\,576}\right) \left(x^*\right)^3 + \ldots \qquad \label{xI4}
\end{IEEEeqnarray} \\
\subsection[Anomalous Dimensions]{Anomalous Dimensions \label{AnomalousDimensionsI}}
Just as for the case of strings on $\text{S}^2$, the anomalous scaling dimensions $\gamma = \mathcal{E} - \mathcal{S}$ of the AdS$_3$ string may be expressed as a function of $\mathcal{S}$:
\begin{IEEEeqnarray}{ll}
\mathcal{E} - \mathcal{S} = \sum_{n = 0}^{\infty} x^n \left(f_n \, \ln x + g_n\right) = \sum_{n = 0}^{\infty} x^n \left[A_n + f_n \, \ln \frac{x}{x_0}\right]\,, \ & \mathcal{E} \equiv \frac{\pi E}{2\sqrt{\lambda}}\,, \ \mathcal{S} \equiv \frac{\pi S}{2\sqrt{\lambda}}, \qquad \label{AnomalousDimensionsI3}
\end{IEEEeqnarray}
where,
\begin{IEEEeqnarray}{c}
x_0 \equiv \exp\left[\frac{\mathcal{S} - b_0}{c_0} + \frac{c_1}{c_0^2}\right] = 16 \, e^{4\mathcal{S} + 3/2}
\end{IEEEeqnarray}
and we also define
\begin{IEEEeqnarray}{l}
f_n \equiv -c_n - \sum_{k = 0}^n \frac{(2k - 3)!!}{\left(2k\right)!!} \cdot d_{n - k} \nonumber \\
g_n \equiv -b_n - \frac{\left(2n - 1\right)!!}{\left(2n + 2\right)!!} - \sum_{k = 0}^n \frac{(2k - 3)!!}{\left(2k\right)!!} \cdot h_{n - k} \,, \quad n = 0,1,2,\ldots \qquad
\end{IEEEeqnarray}
The coefficients $A_n$ are defined as:
\begin{IEEEeqnarray}{ll}
A_n \equiv g_n + f_n \ln x_0 = g_n + f_n \, \left(4\ln 2 + 4\mathcal{S} + \frac{3}{2}\right).
\end{IEEEeqnarray}
\indent For later purposes, it will be useful to obtain the expression of the anomalous dimensions $\gamma = \mathcal{E} - \mathcal{S}$ in terms of $x^*$. We first insert \eqref{xEquationI1} into \eqref{AnomalousDimensionsI3}:
\begin{IEEEeqnarray}{ll}
\mathcal{E} - \mathcal{S} = \frac{\text{a}_0\,f_0}{x} + A_0 + \sum_{n = 1}^{\infty} x^n \left[A_n + \text{a}_0 \, f_{n + 1} + \sum_{k = 0}^{n - 1} f_{n - k - 1} \, \text{a}_{k + 1}\right]. \qquad \label{AnomalousDimensionsI4}
\end{IEEEeqnarray}
Subsequently we plug series \eqref{xI3} and \eqref{xI4} into \eqref{AnomalousDimensionsI4}, which leads us to the following result (for simplicity, we omit higher-order terms):
\begin{IEEEeqnarray}{ll}
\mathcal{E} &- \mathcal{S} = \frac{2}{x^*} - \left(2\,\mathcal{S} + \frac{5}{4}\right) - \frac{9\,x^*}{32} - \left(\frac{\mathcal{S}}{32} + \frac{35}{256}\right) \left(x^*\right)^2 - \left(\frac{5\,\mathcal{S}}{128} + \frac{2.213}{24.576}\right) \left(x^*\right)^3 - \nonumber \\[12pt]
& - \left(\frac{\mathcal{S}^2}{256} + \frac{361 \, \mathcal{S}}{8.192} + \frac{6.665}{98.304}\right) \left(x^*\right)^4 - \left(\frac{19\, \mathcal{S}^2}{2.048} + \frac{1.579 \, \mathcal{S}}{32.768} + \frac{433.501}{7.864.320}\right) \left(x^*\right)^5 + \ldots. \qquad \label{AnomalousDimensionsI5}
\end{IEEEeqnarray}
\indent At this point, a note analogous to that made for strings rotating in $\mathbb{R}\times\text{S}^2$, concerning the structure of the large-spin expansion of the inverse spin function and the anomalous dimensions of strings spinning in AdS$_3$, should be made. Consider the following kinds of terms ($n = 0,1,2,\ldots$):
\begin{IEEEeqnarray}{l}
\text{Leading Terms (L): } \frac{\ln^n\mathcal{S}}{\mathcal{S}^n} \nonumber \\
\text{Next-to-Leading/Subleading Terms (NL): } \frac{\ln^n\mathcal{S}}{\mathcal{S}^{n+1}} \nonumber \\
\text{NNL Terms: } \frac{\ln^n\mathcal{S}}{\mathcal{S}^{n+2}} \nonumber \\
\hspace{2cm} \vdots
\end{IEEEeqnarray}
\newpage In what follows we shall find that, in the expansion of the inverse spin function $x = x\left(\mathcal{S}\right)$, all the leading terms $\ln^n\mathcal{S} / \mathcal{S}^n$ are absent whereas, in the anomalous dimensions $\gamma = \mathcal{E} - \mathcal{S}$, all of the above terms are present. Therefore the large-spin expansion of $\gamma$ assumes the following form:
\begin{IEEEeqnarray}{ll}
\mathcal{E} - \mathcal{S} = \rho_c \, \ln\mathcal{S} &+ \sum_{n = 0}^{\infty}\sum_{k = 0}^{n} \rho_{(nk)} \, \frac{\ln^{k}\mathcal{S}}{\mathcal{S}^n} = \rho_c \, \ln\mathcal{S} + \rho_0 + \sum_{n = 1}^{\infty}\rho_{(nn)}\frac{\ln^n\mathcal{S}}{\mathcal{S}^n} + \sum_{n = 2}^{\infty}\rho_{(nn-1)}\frac{\ln^{n-1}\mathcal{S}}{\mathcal{S}^{n}} + \nonumber \\[6pt]
& + \sum_{n = 3}^{\infty}\rho_{(nn-2)}\frac{\ln^{n-2}\mathcal{S}}{\mathcal{S}^{n}} + \ldots + \frac{\rho_{1}}{\mathcal{S}} + \frac{\rho_{2}}{\mathcal{S}^2} + \frac{\rho_{3}}{\mathcal{S}^3} + \ldots \label{AnomalousDimensionsI6}
\end{IEEEeqnarray}
Note also the presence of the exceptional (super-) leading term $f \ln\mathcal{S}$. Equivalently we may write:
\begin{IEEEeqnarray}{ll}
E - S = &f \, \ln\left(S/\sqrt{\lambda}\right) + \sum_{n = 0}^{\infty}\sum_{k = 0}^{n} f_{(nk)} \, \frac{\ln^{k}\left(S/\sqrt{\lambda}\right)}{S^n} = f \, \ln\left(S/\sqrt{\lambda}\right) + f_0 + \nonumber \\[12pt]
& + \sum_{n = 1}^{\infty}f_{(nn)}\frac{\ln^n\left(S/\sqrt{\lambda}\right)}{S^n} + \sum_{n = 2}^{\infty} f_{(nn-1)} \frac{\ln^{n-1}\left(S/\sqrt{\lambda}\right)}{S^{n}} + \sum_{n = 3}^{\infty} f_{(nn-2)} \frac{\ln^{n-2}\left(S/\sqrt{\lambda}\right)}{S^{n}} + \nonumber \\[12pt]
&+ \ldots + \frac{f_{1}}{S} + \frac{f_{2}}{S^2} + \frac{f_{3}}{S^3} + \ldots, \qquad \label{AnomalousDimensionsI7}
\end{IEEEeqnarray}
where
\begin{IEEEeqnarray}{c}
\rho_c = \frac{\pi\,f}{2\sqrt{\lambda}} \,, \quad \rho_0 = \frac{\pi}{2\sqrt{\lambda}}\left(f_0 + f\ln\frac{2}{\pi}\right) \,, \quad \rho_1 = \left(\frac{\pi}{2\sqrt{\lambda}}\right)^2 \left(f_1 + f_{11}\ln\frac{2}{\pi}\right) \nonumber \\[6pt]
\rho_2 = \left(\frac{\pi}{2\sqrt{\lambda}}\right)^3 \left(f_2 + f_{21}\ln\frac{2}{\pi} + f_{22}\ln^2\frac{2}{\pi}\right) \nonumber \\[6pt]
\rho_3 = \left(\frac{\pi}{2\sqrt{\lambda}}\right)^4 \left(f_3 + f_{31}\ln\frac{2}{\pi} + f_{32}\ln^2\frac{2}{\pi} + f_{33}\ln^3\frac{2}{\pi}\right) \nonumber \\[6pt]
\rho_{(nn)} = \left(\frac{\pi}{2\sqrt{\lambda}}\right)^{n + 1} \cdot f_{(nn)} \,, \quad \rho_{(nn-1)} = \left(\frac{\pi}{2\sqrt{\lambda}}\right)^{n + 1} \left(f_{(nn-1)} + n\,f_{(nn)}\ln\frac{2}{\pi}\right) \nonumber \\[6pt]
\rho_{(nn-2)} = \left(\frac{\pi}{2\sqrt{\lambda}}\right)^{n + 1} \left(f_{(nn-2)} + \left(n-1\right)f_{(nn-1)}\ln\frac{2}{\pi} + \frac{n\left(n-1\right)}{2}\ln\frac{2}{\pi}\right).
\end{IEEEeqnarray} \\
\subsection[Leading Terms]{Leading Terms \label{LeadingI}}
We may now employ formula \eqref{xstar1} in order to calculate the leading in $\mathcal{S}$ terms of series \eqref{AnomalousDimensionsI3}:
\begin{IEEEeqnarray}{c}
x^* = x_0 \cdot e^{W_{-1}\left(\text{a}_0 / x_0\right)} = \frac{\text{a}_0}{W_{-1}\left(\text{a}_0 / x_0\right)} = \frac{-4}{W_{-1}\left[-e^{-4\mathcal{S} - 3/2}/4\right]}. \label{xstar4}
\end{IEEEeqnarray}
Note that, as $\mathcal{S} \rightarrow +\infty$ we should have $x^* \rightarrow 0^+$, behavior that is only possible if $W \rightarrow -\infty$, i.e.\ in the $W_{-1}$ branch of Lambert's function (on the contrary $W_0 \rightarrow 0^-$ for $\mathcal{S} \rightarrow +\infty$, which would cause $x^*$ to blow up as $x^* \rightarrow +\infty$, cf. figure \ref{Graph:LambertFunction}). Using expansion \eqref{LambertSeries-1} of the W-function, we may obtain the inverse spin function $x = x\left(\mathcal{S}\right)$ up to leading order. Let us first calculate $1/x^*$:
\small\begin{IEEEeqnarray}{ll}
\frac{1}{x^*} = \mathcal{S} &+ \frac{\ln\mathcal{S}}{4} + \ln2 + \frac{3}{8} + \sum_{n = 1}^\infty \frac{\left(-1\right)^{n+1}}{4 n} \left(\frac{4\ln2 + 3}{8\,\mathcal{S}}\right)^n - \sum_{n,q = 0}^\infty \sum_{m = 1}^\infty \sum_{p = 0}^{m} \frac{\left(-1\right)^m}{4^{n+m+1} \, m!} \left[\begin{array}{c}n+m\\n+1\end{array}\right] \cdot \nonumber \\[12pt]
&\cdot  \left(\begin{array}{c}-n - m\\q\end{array}\right)\left(\begin{array}{c}m\\p\end{array}\right) \frac{\ln^p\mathcal{S}}{\mathcal{S}^{n + m}} \left(2\ln2 - \sum_{k = 1}^\infty \frac{\left(-1\right)^{k}}{k} \left(\frac{4\ln2 + 3}{8\,\mathcal{S}}\right)^k\right)^{m-p} \left(\frac{4\ln2 + 3}{8\,\mathcal{S}}\right)^q \label{xstar5}
\end{IEEEeqnarray} \normalsize
where, for large $\mathcal{S}$, we have written
\begin{IEEEeqnarray}{l}
\ln\left[4\mathcal{S} + 2 \ln2 + \frac{3}{2}\right] = \ln\mathcal{S} + 2\ln2 + \sum_{n = 1}^\infty \frac{\left(-1\right)^{n+1}}{n} \left(\frac{4\ln2 + 3}{8\,\mathcal{S}}\right)^n
\end{IEEEeqnarray}
and the unsigned Stirling numbers of the first kind $\left[\begin{array}{c}n + m \\ n + 1\end{array}\right]$ are defined in appendix \ref{LambertAppendix}. To obtain $x^*$, we expand the inverse of series \eqref{xstar5}:
\small\begin{IEEEeqnarray}{ll}
x^* =\frac{1}{\mathcal{S}} \cdot \Bigg\{1 &+ \frac{\ln\mathcal{S}}{4\,\mathcal{S}} + \left(\ln2 + \frac{3}{8}\right)\frac{1}{\mathcal{S}} + 2\,\sum_{n = 1}^\infty \frac{\left(-1\right)^{n+1}}{n} \frac{\left(4\ln2 + 3\right)^{n}}{8\,\mathcal{S}^{n + 1}} - \sum_{n,q = 0}^\infty \sum_{m = 1}^\infty \sum_{p = 0}^{m} \frac{\left(-1\right)^m}{4^{n+m+1} \, m!} \cdot \nonumber \\[12pt]
& \cdot \left[\begin{array}{c}n+m\\n+1\end{array}\right]\left(\begin{array}{c}-n - m\\q\end{array}\right) \left(\begin{array}{c}m\\p\end{array}\right) \frac{\ln^p\mathcal{S}}{\mathcal{S}^{n + m + 1}} \left(2\ln2 - \sum_{k = 1}^\infty \frac{\left(-1\right)^{k}}{k} \left(\frac{4\ln2 + 3}{8\,\mathcal{S}}\right)^k\right)^{m-p} \cdot \nonumber \\[12pt]
& \cdot\left(\frac{4\ln2 + 3}{8\,\mathcal{S}}\right)^q\Bigg\}^{-1}. \label{xstar6}
\end{IEEEeqnarray} \normalsize
\indent A few comments are in order, before proceeding to the calculation of the corresponding series for the anomalous dimensions. First observe that series \eqref{xstar5} for $1/x^*$ contains all kinds of small terms (i.e.\ terms $\rightarrow0$ as $\mathcal{S}\rightarrow\infty$): leading terms $\ln^n\mathcal{S}/\mathcal{S}^n$, subleading terms $\ln^n\mathcal{S}/\mathcal{S}^{n+1}$, next-to-subleading terms $\ln^n\mathcal{S}/\mathcal{S}^{n+2}$, etc.\ up to $1/\mathcal{S}^n$ terms ($n = 1,\,2,\,3\,\ldots$). Series \eqref{xstar6} for $x^*$ on the contrary, does not contain leading terms. It is to be expected that $\left(x^*\right)^2$ will not contain leading and subleading terms, in $\left(x^*\right)^3$ leading, subleading and next-to-subleading terms will be absent, etc. Therefore, to calculate the $E - S$ series up to leading order in $\mathcal{S}$, we need no more than the first two terms of \eqref{AnomalousDimensionsI5},
\begin{IEEEeqnarray}{ll}
\mathcal{E} - \mathcal{S} \Big|_{(\text{L} + \ldots)} = \frac{2}{x^*} - \left(2\,\mathcal{S} + \frac{5}{4}\right),
\end{IEEEeqnarray}
since the other terms of series \eqref{AnomalousDimensionsI5} only contribute to NL orders. We obtain, \\
\small\begin{IEEEeqnarray}{ll}
\mathcal{E} - \mathcal{S}\Big|_{(\text{L} + \ldots)} = \frac{1}{2} &\ln\mathcal{S} + \left(2\ln2 - \frac{1}{2}\right) + \sum_{n = 1}^\infty \frac{\left(-1\right)^{n+1}}{2 n} \left(\frac{4\ln2 + 3}{8\,\mathcal{S}}\right)^n - \frac{1}{2} \sum_{n,q = 0}^\infty \sum_{m = 1}^\infty \sum_{p = 0}^{m} \frac{\left(-1\right)^m}{4^{n+m}} \cdot \nonumber \\[12pt]
& \cdot \frac{1}{m!}\left[\begin{array}{c}n+m\\n+1\end{array}\right] \left(\begin{array}{c}-n - m\\q\end{array}\right) \left(\begin{array}{c}m\\p\end{array}\right) \frac{\ln^p\mathcal{S}}{\mathcal{S}^{n + m}} \left(\frac{4\ln2 + 3}{8\,\mathcal{S}}\right)^q \Big[2\ln2 - \sum_{k = 1}^\infty \frac{\left(-1\right)^{k}}{k} \cdot \nonumber \\[12pt]
& \cdot \left(\frac{4\ln2 + 3}{8\,\mathcal{S}}\right)^k\Big]^{m-p}. \qquad
\end{IEEEeqnarray} \normalsize \\
For $p = m$ and $n = q = 0$ we read off the coefficients of the leading terms:
\begin{IEEEeqnarray}{l}
\rho_{(mm)} = - \frac{1}{2} \frac{\left(-1\right)^m}{4^m \, m!} \cdot \left[\begin{array}{c}m\\1\end{array}\right] \left(\begin{array}{c}- m\\0\end{array}\right) \left(\begin{array}{c}m\\m\end{array}\right) = \frac{\left(-1\right)^{m + 1}}{4^m} \frac{1}{2m}, \label{GKPILeading1}
\end{IEEEeqnarray}
which agrees with the results of \cite{GeorgiouSavvidy11} (we have used the first property of unsigned Stirling numbers in \eqref{StirlingNumbers2}). Also,
\begin{IEEEeqnarray}{l}
\rho_{c} = \frac{1}{2} \quad \& \quad \rho_0 = 2\ln2 - \frac{1}{2}. \label{GKPILeading2}
\end{IEEEeqnarray} \\
\subsection[Next-to-Leading Terms]{Next-to-Leading Terms \label{SubleadingI}}
To extract the subleading coefficients, we have to keep the following terms of series \eqref{AnomalousDimensionsI5}
\begin{IEEEeqnarray}{c}
\mathcal{E} - \mathcal{S}\Big|_{(\text{L} + \text{NL} + \ldots)} = \frac{2}{x^*} - \left(2\,\mathcal{S} + \frac{5}{4}\right) - \frac{9\,x^*}{32} - \frac{\mathcal{S}}{32}\left(x^*\right)^2 \qquad
\end{IEEEeqnarray}
and read off the terms that contribute to subleading order from equations \eqref{xstar5}-\eqref{xstar6}. The result is (with the aid of \eqref{StirlingNumbers2}):
\begin{IEEEeqnarray}{l}
\rho_{(m+1,m)} = \frac{1}{2} \frac{\left(-1\right)^{m + 1}}{4^{m + 1}} \left[H_m + \frac{m}{4}+ 1 - 4\ln2\right], \label{GKPISubleading1}
\end{IEEEeqnarray}
in conformity with \cite{GeorgiouSavvidy11}. We also confirm the coefficient $\rho_1$ in expansion \eqref{AnomalousDimensionsI6}:
\begin{IEEEeqnarray}{l}
\rho_{1} = \frac{\ln 2}{2} - \frac{1}{8}. \label{GKPISubleading2}
\end{IEEEeqnarray}

\subsection[NNL Terms]{NNL Terms \label{NexttosubleadingI}}
We may also go to higher subleading orders, e.g. the NNL one. In this case the terms of \eqref{AnomalousDimensionsI5} that contribute are:
\begin{IEEEeqnarray}{ll}
\mathcal{E} - \mathcal{S}\Big|_{\text{L} + \text{NL} + \text{NNL} + \ldots} = \frac{2}{x^*} &- \left(2\,\mathcal{S} + \frac{5}{4}\right) - \frac{9\,x^*}{32} - \left(\frac{\mathcal{S}}{32} + \frac{35}{256}\right) \left(x^*\right)^2 - \frac{5\,\mathcal{S}}{128} \left(x^*\right)^3 \nonumber \\[6pt]
& - \frac{\mathcal{S}^2}{256} \left(x^*\right)^4. \qquad
\end{IEEEeqnarray} \normalsize
We now can read off all the NNL terms (using property \eqref{StirlingNumbers2} of unsigned Stirling numbers):
\begin{IEEEeqnarray}{ll}
\rho_{(m+2,m)} = \frac{\left(-1\right)^{m + 1}}{4^{m+3}} \cdot \left(m + 1\right) \cdot \Bigg\{H^2_{m + 1} &- H^{(2)}_{m + 1} + \frac{1}{2} \left(m - 16\ln 2 + 5\right) \cdot H_{m + 1} + \frac{m^2}{24} - \nonumber \\[12pt]
& - \left(2\ln 2 + \frac{1}{24}\right) m + 16\ln^2 2 - 10 \ln2\Bigg\}. \label{GKPINexttoSubleading1}
\end{IEEEeqnarray}
For the coefficient $\rho_2$ of \eqref{AnomalousDimensionsI6} we find:
\begin{IEEEeqnarray}{l}
\rho_{2} = -\frac{\ln^2 2}{4} + \frac{9\ln2}{32} - \frac{5}{128}. \label{GKPINexttoSubleading2}
\end{IEEEeqnarray}
The above results agree completely with the (first three) terms of the series \eqref{AnomalousDimensionsI7} that were computed in \cite{Beccariaetal09, BeccariaForiniMacorini10}. \\
\subsection[Higher-Order Terms]{Higher-Order Terms \label{HigherOrderTermsI}}
Similarly, we may get going to higher and higher orders. Using Mathematica we find e.g.\ (cf. \eqref{MathematicaAnomalousDimensionsI1}),
\begin{IEEEeqnarray}{l}
\rho_{3} = \frac{\ln^3 2}{6} - \frac{3\ln^2 2}{8} + \frac{11\ln2}{64} - \frac{7}{384} \nonumber \\[12pt]
\rho_{4} = - \frac{\ln^4 2}{8} + \frac{43\ln^3 2}{96} - \frac{51\ln^2 2}{128} + \frac{937\ln2}{8192} - \frac{1919}{196.608} \nonumber \\[12pt]
\rho_{5} = \frac{\ln^5 2}{10} - \frac{49\ln^4 2}{96} + \frac{23\ln^3 2}{32} - \frac{777\ln^2 2}{2048} + \frac{1963\ln2}{24.576} - \frac{7423}{1.310.720} \nonumber \\[12pt]
\rho_{6} = - \frac{\ln^6 2}{12} + \frac{17\ln^5 2}{30} - \frac{581\ln^4 2}{512} + \frac{11.401\ln^3 2}{12.288} - \frac{67.715\ln^2 2}{196.608} + \frac{30.085\ln2}{524.288} - \frac{218.431}{62.914.560}. \nonumber
\end{IEEEeqnarray} \\
\indent We may also express formulas \eqref{xI3}-\eqref{AnomalousDimensionsI5} for the inverse spin function $x\left(\mathcal{S}\right)$ and the anomalous dimensions $\gamma = \mathcal{E} - \mathcal{J}$ of the long operators Tr$\left[\mathcal{Z} \, \mathcal{D}_+^S \, \mathcal{Z}\right] + \ldots$ of $\mathcal{N} = 4$ SYM at strong 't Hooft coupling in terms of Lambert's W-function. Plugging equation \eqref{xstar4} into \eqref{xI3} and \eqref{AnomalousDimensionsI5} we find, respectively (for simplicity, only the first few terms are displayed):
\begin{IEEEeqnarray}{ll}
x = & - \frac{4}{W_{-1}} - \frac{8\,\mathcal{S} + 3}{\left(W_{-1}\right)^3} + \bigg[16\mathcal{S} + \frac{23}{4}\bigg]\frac{1}{\left(W_{-1}\right)^4} - \bigg[32\,S^2 + 55\,S + \frac{349}{24}\bigg]\frac{1}{\left(W_{-1}\right)^5} + \bigg[152\,S^2 + \nonumber \\[6pt]
& + \frac{711\,S}{4} + \frac{3745}{96}\bigg]\frac{1}{\left(W_{-1}\right)^6} - \bigg[160\,S^3 + 704\,S^2 + \frac{4765\,S}{8} + \frac{26.659}{240}\bigg]\frac{1}{\left(W_{-1}\right)^7} + \nonumber \\[6pt]
& + \bigg[\frac{3728\,S^3}{3}+\frac{6077\,S^2}{2} + \frac{48.955\,S}{24} + \frac{2.543.083}{7680}\bigg]\frac{1}{\left(W_{-1}\right)^8} - \ldots, \label{xI5}
\end{IEEEeqnarray}
\begin{IEEEeqnarray}{ll}
\mathcal{E} - \mathcal{S} = & - \frac{W_{-1}}{2} - \left(2\,\mathcal{S} + \frac{5}{4}\right) + \frac{9}{8\,W_{-1}} - \bigg[\frac{\mathcal{S}}{2} + \frac{35}{16}\bigg]\frac{1}{\left(W_{-1}\right)^2} + \bigg[\frac{5\,\mathcal{S}}{2} + \frac{2213}{384}\bigg]\frac{1}{\left(W_{-1}\right)^3} - \nonumber \\[6pt]
& - \bigg[\mathcal{S}^2 + \frac{361\,\mathcal{S}}{32} + \frac{6665}{384}\bigg]\frac{1}{\left(W_{-1}\right)^4} + \bigg[\frac{19\,\mathcal{S}^2}{2} + \frac{1579\,\mathcal{S}}{32} + \frac{433501}{7680}\bigg]\frac{1}{\left(W_{-1}\right)^5} - \nonumber \\[6pt]
& - \bigg[\frac{10\,\mathcal{S}^3}{3} + \frac{259\,\mathcal{S}^2}{4} + \frac{81.799\,\mathcal{S}}{384} + \frac{2.963.887}{15.360}\bigg]\frac{1}{\left(W_{-1}\right)^6} + \bigg[\frac{136\,\mathcal{S}^3}{3} + \frac{3069\,\mathcal{S}^2}{8} + \nonumber \\[6pt]
& + \frac{175.481\,\mathcal{S}}{192} + \frac{2.350.780.111}{3.440.640}\bigg]\frac{1}{\left(W_{-1}\right)^7} - \ldots, \label{AnomalousDimensionsI8}
\end{IEEEeqnarray}
where the arguments of the W-functions are $W_{-1}\left(-e^{- 4\mathcal{S} - 3/2}/4\right)$ and $\mathcal{E} \equiv \pi E / 2\sqrt{\lambda}$, $\mathcal{S} \equiv \pi S / 2\sqrt{\lambda}$. Just as for strings in $\mathbb{R}\times\text{S}^2$, the terms of both series \eqref{xI5} and \eqref{AnomalousDimensionsI8} are organized in decreasing order of importance, e.g. in \eqref{AnomalousDimensionsI8} the first two terms contain all the leading coefficients, the first four terms all of the subleading coefficients, and so on. Our results can be verified with a direct calculation in Mathematica (see \eqref{Exact_Value_of_xI1}, \eqref{MathematicaAnomalousDimensionsI1}). \\
\section[Discussion]{Discussion\label{Discussion}}
\begin{figure}
\centering
\includegraphics[scale=0.4]{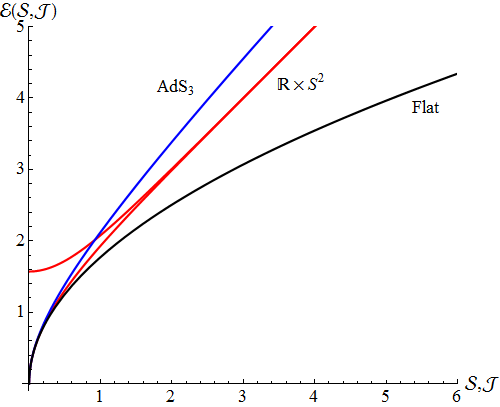}
\caption{$\mathcal{E} = \mathcal{E}\left(\mathcal{S},\mathcal{J}\right)$ for GKP strings in AdS$_3$, S$^2$ and flat spacetimes.} \label{Graph:Comparisons}
\end{figure}

We have computed the large-spin expansion of anomalous dimensions of two single-trace operators of $\mathcal{N} = 4$ super Yang-Mills theory at strong coupling, namely
\begin{IEEEeqnarray}{c}
\mathcal{O}_J = \text{Tr}\left[\Phi\mathcal{Z}^m \, \Phi \, \mathcal{Z}^{J-m}\right] + \ldots \qquad \& \qquad \mathcal{O}_S = Tr\left[\mathcal{Z} \, \mathcal{D}_+^S \, \mathcal{Z}\right] + \ldots, \qquad \label{Operators2}
\end{IEEEeqnarray}
where $\Phi$ is any scalar field of $\mathcal{N} = 4$ SYM that is not used to build $\mathcal{Z}$. According to the AdS/CFT correspondence, the anomalous dimensions of the operators under consideration are given by the dispersion relation of their dual $\text{AdS}_5\times\text{S}^5$ strings. Operator $\mathcal{O}_J$ is dual to single-spin strings that rotate inside $\mathbb{R}\times\text{S}^2$, while $\mathcal{O}_S$ is dual to single-spin strings that rotate inside $\text{AdS}_3$. We have expressed the anomalous dimensions $\gamma$ of the above (long) gauge-theory operators in terms of Lambert's W-function. This results in more compact expressions for $\gamma$'s and simplify the derivation of the corresponding series. We have also found a duality between short and long strings. For each solution of energy $E$ and spin $J$, there exists a dual solution whose energy $E'$ and spin $J'$ are related to the original by equations \eqref{Short-LongII1}--\eqref{Short-LongII2}. These relations are purely classical ($\lambda \gg 1$) but it would be interesting to see if and how they can be promoted to the quantum level. \\
\indent The inversion of elliptic integrals and Jacobi elliptic functions w.r.t.\ the \textit{parameter} $m$, is an active field of research in computational mathematics.\footnote{See e.g. \cite{Fukushima12b}.} It seems though that due to the presence of the logarithmic singularity at $m = 1$ (cf. appendix \ref{EllipticFunctionsAppendix}), no significant progress in actually calculating these inverses has yet been made. In our case, we have noticed that equation \eqref{xEquationII1} may be solved by the Lagrange-B\"{u}rmann formula and that the result may then be expressed in terms of Lambert's W-function. In the case of AdS$_3$ a slightly different process had to be followed, as the equation to be inverted \eqref{xEquationI1} is more difficult, because of the additional $1/x$ term on the r.h.s.\footnote{This difficulty partly accounts for the somewhat unconventional organization of our paper, discussing the GKP configuration (II) before (I).} The presence of this $1/x$ term, in fact leads to the appearance of the branch $W_{-1}$ of Lambert's function instead of $W_0$ and subsequently to logarithmic instead of exponential corrections for the inverse spin function and the anomalous dimensions. \\
\indent Equations \eqref{InverseSpinFunctionII1}--\eqref{AnomalousDimensions16} and \eqref{xI5}--\eqref{AnomalousDimensionsI8} that we have found, are only valid up to a certain subleading order. It would be interesting if generalizations to all subleading orders via a general formula or a recursive process could be found. As we have already said, we believe that the Lambert functions will keep appearing to all subleading orders. One could also consider changing branches in Lambert's W-functions. Changing between the Lambert branches $W_0$ and $W_{-1}$ means that the inverse spin function either blows up (i.e.\ $x\rightarrow\pm\infty$) or in general does not have the desired behavior $x\rightarrow 0$. Conversely, we saw that a single sign flip in the argument of Lambert's function (cf. \eqref{AnomalousDimensions16}, \eqref{GKPIICircularAnomalousDimensions2}) carries us from folded ($\omega>1$) to circular ($\omega<1$) strings on S$^2$ and vice-versa. This could be the manifestation of a more profound link between the two domains. It is also possible that the Lambert function formalism could help make apparent the symmetries that are hidden inside the large-spin expansions (e.g. one such symmetry could be the \textit{near conjugate symmetry} $W_k\left(\overline{z}\right) = \overline{W}_{-k}\left(z\right)$). \\
\indent Our results from the study of GKP strings on AdS \eqref{AnomalousDimensionsI8} point out that the anomalous dimensions of long twist-2 operators of $\mathcal{N} = 4$ SYM theory at strong coupling may be expressed in terms of Lambert's W-function. This could also be the case at weak coupling. For the cousin theory of QCD, long twist operators (responsible for scaling violations in DIS) could also admit an analogous Lambert parametrization. Note that the exact 3-loop running coupling constant of QCD has already been found to be expressible in terms of Lambert's W-function \cite{KhuriRen88, AppelquistRatnaweeraTerningWijewardhana98, GardiGrunbergKarliner98, Magradze99}:
\begin{IEEEeqnarray}{c}
\alpha_s\left(Q^2\right) = \frac{- \pi/c}{1 - c_2/c^2 + W\left(z\right)}, \label{QCDRunningCoupling}
\end{IEEEeqnarray}
where $c_2$ is a renormalization scheme-dependent constant and
\small\begin{IEEEeqnarray}{c}
\beta_0 \equiv \frac{1}{4}\left(11 - \frac{2}{3}n_f\right), \quad c \equiv \frac{1}{4\beta_0}\left[102 - \frac{38}{3}n_f\right], \quad
z \equiv -\frac{1}{c}\exp\left[-1 + \frac{c_2}{c^2} - \frac{\beta_0\,t}{c}\right], \quad t \equiv \ln\left(\frac{Q^2}{\Lambda^2}\right). \nonumber
\end{IEEEeqnarray}\normalsize
It would be surprising if the anomalous dimensions of long, twist QCD operators didn't have anything to do with W (at least in strong coupling).\footnote{Note that the branch of the W-function in the QCD case \eqref{QCDRunningCoupling} depends on the number of flavors $n_f$. For $c>0 \Leftrightarrow z<0$ the relevant branch is $W_{-1}$, while for $c<0 \Leftrightarrow z>0$ the branch is $W_0$ \cite{GardiGrunbergKarliner98}.} Thermal backgrounds, as well as dilaton geometries within holographic frameworks \cite{CsakiReece06} could also be susceptible to analogous W-formulations since, in the language of holography, Einstein's equations are nothing more than RG equations and these have also been solved in terms of Lambert's W-function \cite{Sonoda13b, CurtrightZachos10b}. Our analysis suggests that a generic feature of the solution of the RG equations at any loop-order is that they can be expressed in terms of Lambert's W-function (to see this compare equation \eqref{xEquationI1} with the integral of the generic RG-equation $\beta(x) = \mu^2 dx /d\mu^2 = - x^2\,\sum \beta_n \, x^n$.). \\
\indent All of our expressions for long/fast strings have been verified with Mathematica (see appendix \ref{Long-Fast-StringsAppendix}). We have also considered short strings (appendix \ref{Short-Slow-StringsAppendix}). Since the elliptic integrals do not have a logarithmic singularity for eccentricities smaller than one, $\mathcal{E} = \mathcal{E}\left(\mathcal{J}\right)$ and $\mathcal{E} = \mathcal{E}\left(\mathcal{S}\right)$ can be obtained by simple series reversion in Mathematica. It is possible that these short series also afford an analysis into sums of W-functions. Compact forms for the short series would be rather interesting because of the relationship that short spinning strings bear to closed strings rotating in flat spacetimes. In this case $\mathcal{E} = \sqrt{\pi\,\mathcal{S}}$, which is just the first term of the short series (either in $\mathbb{R}\times\text{S}^2$ or AdS$_3$). In figure \ref{Graph:Comparisons} we have plotted in a common diagram the energy as a function of spin, for all these three cases of folded closed GKP strings: \eqref{GKPIICircularEnergy1}--\eqref{GKPIICircularSpin1} and \eqref{GKPIIEnergy2}--\eqref{GKPIISpin2} for strings on $\mathbb{R}\times\text{S}^2$, \eqref{GKPIEnergy2}--\eqref{GKPISpin2} for strings in AdS$_3$ and $\mathcal{E} = \sqrt{\pi\,\mathcal{S}}$ for strings rotating in flat spacetime. \\
\indent It is known that both GKP strings (I) and (II) that we have studied in our paper, can be formed by the superposition of other elementary string-excitations, spiky strings and giant magnons (GMs) respectively. It would be then natural to inquire whether our findings have any consequences whatsoever to the computation of the general dispersion relations of these elementary excitations. \\
\indent As we have already mentioned in the introduction, strings on $\mathbb{R}\times\text{S}^2$ are the sum of two giant magnons of maximum momentum $p = \pi$ and angular momentum $J/2$. The all-loop, infinite-volume Beisert dispersion relation \eqref{GiantMagnon2} \cite{Beisert05b}, converges to the Hofman-Maldacena \cite{HofmanMaldacena06} relation \eqref{GiantMagnon1} at strong coupling and receives finite-size corrections as the volume is gradually decreased. Arutyunov, Frolov and Zamaklar (AFZ) \cite{ArutyunovFrolovZamaklar06} and Astolfi, Forini, Grignani and Semenoff \cite{AstolfiForiniGrignaniSemenoff07} have calculated the finite-size corrections to the giant magnon \eqref{GiantMagnon3}, while Klose and McLoughlin \cite{KloseMcLoughlin08} have provided the leading terms \eqref{GiantMagnon4}. \\
\indent Our result \eqref{AnomalousDimensions16} completely agrees with formula (3.20) of AFZ \cite{ArutyunovFrolovZamaklar06}.\footnote{In comparing our results to those of AFZ one should note a difference in the definition of $\mathcal{J} \equiv \pi J/2\sqrt{\lambda}$, given in AFZ as $\mathcal{J}_{\left\{\text{AFZ}\right\}} \equiv 2\pi J/\sqrt{\lambda}$.} We have also found that the finite-size corrections of giant magnons can be expressed in terms of Lambert's function $W\left(- 16 \mathcal{J}^2 \cot^2\left(p/2\right) e^{- 2\mathcal{J}\csc p/2 - 2}\right)$ as follows:\footnote{Our result is valid in the conformal gauge of the string Polyakov action ($\gamma_{ab} = \eta_{ab}$) and the static time gauge, $t =\tau$. The details of the derivation will be given in a forthcoming publication.}
\begin{IEEEeqnarray}{ll}
\mathcal{E} - \mathcal{J} = \sin\frac{p}{2} &+ \frac{1}{4\mathcal{J}^2}\tan^2\frac{p}{2}\sin^3\frac{p}{2}\left[W + \frac{W^2}{2}\right] - \frac{1}{16\mathcal{J}^3}\tan^4\frac{p}{2}\sin^2\frac{p}{2}\bigg[\left(3\cos p + 2\right)W^2 + \nonumber \\[6pt]
& + \frac{1}{6}\left(5\cos p + 11\right)W^3\bigg] - \frac{1}{512\mathcal{J}^4}\tan^6\frac{p}{2}\sin\frac{p}{2}\Bigg\{\left(7\cos p - 3\right)^2\frac{W^2}{1 + W} - \nonumber \\[6pt]
& - \frac{1}{2}\left(25\cos2p -188\cos p -13\right)W^2 - \frac{1}{2}\left(47\cos2p + 196\cos p - 19\right)W^3 - \nonumber \\[6pt]
& - \frac{1}{3}\left(13\cos2p + 90\cos p + 137\right)W^4\Bigg\} + \ldots \qquad \label{GiantMagnon5}
\end{IEEEeqnarray}
Upon expanding \eqref{GiantMagnon5} we recover formulas (5.14) of AFZ \cite{ArutyunovFrolovZamaklar06} and (39) of Astolfi-Forini-Grignani-Semenoff \cite{AstolfiForiniGrignaniSemenoff07}. Especially for the above formula, we have set $\mathcal{E} \equiv \pi E/\sqrt{\lambda}$ and $\mathcal{J} \equiv \pi J/\sqrt{\lambda}$. The Klose-McLoughlin series \eqref{GiantMagnon4} is retrieved by letting $L_{\text{eff}} = 2\mathcal{J}\csc p/2$:
\begin{IEEEeqnarray}{c}
E - J = \frac{\sqrt{\lambda}}{\pi} \, \sin\frac{p}{2} \, \Bigg\{1 + L_{\text{eff}}^{-2}\,\tan^2\frac{p}{2}\left(W + \frac{1}{2}W^2\right)\Bigg\}, \qquad
\end{IEEEeqnarray}
so that the arguments of the W-functions are $W\left(-4\,L_{\text{eff}}^2\,\cos^2\left(p/2\right)\,e^{-L_{\text{eff}}}\right)$. We should note in passing that the L\"{u}scher corrections that were first calculated in \cite{JanikLukowski07} (equation (6) ibid.), completely agree with AFZ and therefore our results agree with \cite{JanikLukowski07} too. One could also consider further extending these findings to the GMs of ABJM Theory. \\
\indent Similar considerations should also apply to spiky strings \cite{Kruczenski05}, since strings on AdS$_3$ can be thought of as 2-spike Kruczenski strings. One may keep thinking more and more applications of the Lambert formalism. For example the expression of finite-size corrections to the energy of GMs in $\gamma$-deformed backgrounds\footnote{Aka real Lunin-Maldacena backgrounds.} \cite{ChuGeorgiouKhoze06} is very reminiscent of the undeformed ones \eqref{GiantMagnon3}:
\begin{IEEEeqnarray}{ll}
E - J = &\frac{\sqrt{\lambda}}{\pi} \, \sin\frac{p}{2} \, \Bigg\{1 - 4\,\sin^2\frac{p}{2}\,\cos\Xi\,e^{- 2 - 2\pi J/\sqrt{\lambda} \sin\frac{p}{2}} + \ldots\Bigg\}\,, \quad \Xi \equiv \frac{2\pi\left(n_2 - \beta\,J\right)}{2^{3/2}\cos^3 p/4}, \qquad
\end{IEEEeqnarray}
where $n_2$ is the integer string winding number and $\beta$ is the real deformation parameter, satisfying $\left|n_2 - \beta\,J\right|\leq 1/2$ \cite{BykovFrolov08}. For more applications in this direction, one could consult the review article \cite{Zoubos10}. Finally, finite-size effects may be studied for higher dimensional analogues of GMs and single spike strings, e.g. for M2-branes on $\text{AdS}_4\times\text{S}^7$ \cite{AhnBozhilov08b}. Our results are also directly applicable to stringy membranes that rotate inside AdS$_{4/7}\times\text{S}^{7/4}$ \cite{HartnollNunez02, AxenidesFloratosLinardopoulos13a}. \\
\acknowledgments
We would like to thank M. Axenides, D. Giataganas, S. Nicolis and G. Savvidy for many illuminating discussions. \\
\indent E.F. kindly acknowledges the hospitality of CERN Theory Group. G.L. kindly acknowledges the hospitality of APCTP during the program "Holography 2013: Gauge/ gravity duality and strongly correlated systems". \\
\indent The research of E.F. is implemented under the "ARISTEIA" action (Code no.1612, D.654) and title "Holographic Hydrodynamics" of the "operational programme education and lifelong learning" and is co-funded by the European Social Fund (ESF) and National Resources. The research of G.G. and G.L. is supported in part by the General Secretariat for Research and Technology of Greece and from the European Regional Development Fund MIS-448332-ORASY (NSRF 2007--13 ACTION, KRIPIS). \\
\appendix
\section[Lambert's W-Function]{Lambert's W-Function \label{LambertAppendix}}
\begin{figure}
\begin{center}
\includegraphics[scale=0.4]{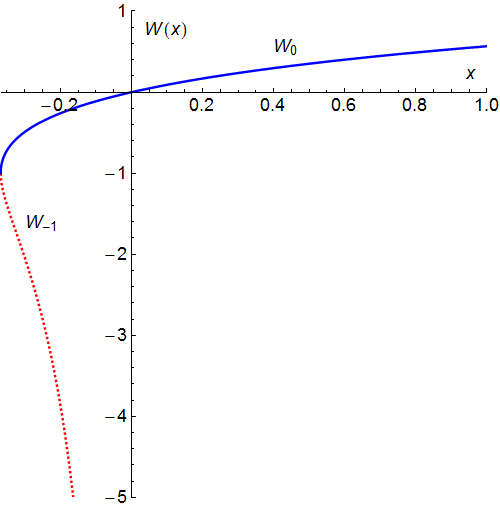}
\caption{Lambert's W-function.} \label{Graph:LambertFunction}
\end{center}
\end{figure}

In this appendix we briefly review Lambert's W-function. It is defined by the following implicit relation:
\begin{IEEEeqnarray}{c}
W\left(z\right)\,e^{W\left(z\right)} = z \Leftrightarrow W\left(z\,e^z\right) = z. \label{LambertDefinition2}
\end{IEEEeqnarray}
$W\left(x\right)$ has two real branches, $W_0\left(x\right)$ in $\left[-e^{-1},\infty\right)$ and $W_{-1}\left(x\right)$ in $\left[-e^{-1},0\right]$, plotted in figure \ref{Graph:LambertFunction}. The branch point is $\left(-e^{-1},-1\right)$. The Taylor series at $x = 0$ for each of the two branches are \cite{CorlessGonnetHareJeffreyKnuth96}:
\begin{IEEEeqnarray}{l}
W_0\left(x\right) = \sum_{n = 0}^\infty \left(-1\right)^n\frac{\left(n+1\right)^n}{\left(n+1\right)!}\cdot x^{n+1} = \sum_{n = 1}^\infty \left(-1\right)^{n + 1} \frac{n^{n - 1}}{n!}\cdot x^n\,, \quad \left|x\right| \leq e^{-1} \label{LambertSeries0} \\[12pt]
W_{-1}\left(x\right) = \ln \left|x\right| - \ln\ln \left|x\right| + \sum_{n = 0}^\infty \sum_{m = 1}^\infty \frac{\left(-1\right)^n}{m!} {n + m \brack n + 1} \left(\ln \left|x\right|\right)^{-n-m} \left(\ln\ln \left|x\right|\right)^m, \qquad \label{LambertSeries-1}
\end{IEEEeqnarray} \\[6pt]
where the unsigned Stirling numbers of the first kind, $\left[\begin{array}{c}n+m\\n+1\end{array}\right]$ can be defined recursively as \cite{Comtet74}:
\begin{IEEEeqnarray}{c}
\left[\begin{array}{c}n \\ k\end{array}\right] = \left[\begin{array}{c}n - 1 \\ k - 1\end{array}\right] + \left(n - 1\right)\left[\begin{array}{c}n - 1 \\ k\end{array}\right] \quad \& \quad \left[\begin{array}{c}n \\ 0\end{array}\right] = \left[\begin{array}{c}0 \\ k\end{array}\right] = 0\,, \ \left[\begin{array}{c}0 \\ 0\end{array}\right] = 1\,, \  n,k\geq 1. \qquad \label{StirlingNumbers1}
\end{IEEEeqnarray}
In our paper we have used the following identities of unsigned Stirling numbers:
\begin{IEEEeqnarray}{c}
\left[\begin{array}{c}n \\ 1\end{array}\right] = \left(n - 1\right)!\,, \quad \left[\begin{array}{c}n \\ 2\end{array}\right] = \left(n - 1\right)!\,H_{n - 1}\,, \quad \left[\begin{array}{c}n \\ 3\end{array}\right] = \frac{1}{2}\left(n - 1\right)!\left[H_{n - 1}^2 - H_{n - 1}^{(2)}\right]. \qquad \label{StirlingNumbers2}
\end{IEEEeqnarray}
The W-function also provides a series for the expression $x^{x^{x^{\ldots}}}$. The result is:
\begin{IEEEeqnarray}{c}
x^{x^{x^{\ldots}}} = \tensor[^\infty]{\left(x^{z}\right)}{} = \frac{W\left(-\ln x\right)}{-\ln x}.
\end{IEEEeqnarray}

\indent Due to its defining property \eqref{LambertDefinition2}, the derivatives and antiderivatives of Lambert's function get significantly simplified. Below, we give a list of some useful expressions involving the $W_0$ function:
\begin{IEEEeqnarray}{c}
W'\left(x\right) = \frac{W\left(x\right)}{x\left(1 + W\left(x\right)\right)} \label{Lambert1} \\[12pt]
x\,W'\left(x\right) = \sum_{n = 1}^\infty \left(-1\right)^{n + 1} \frac{n^n}{n!}\cdot x^n = \frac{W\left(x\right)}{1 + W\left(x\right)} \label{Lambert2} \\[12pt]
x\,\left(x\,W'\left(x\right)\right)' = \sum_{n = 1}^\infty \left(-1\right)^{n + 1} \frac{n^{n + 1}}{n!}\cdot x^n = \frac{W\left(x\right)}{\left(1 + W\left(x\right)\right)^3} \label{Lambert3} \\[12pt]
\int W\left(x\right) \ dx = x\left(W\left(x\right) - 1 + \frac{1}{W\left(x\right)}\right) \label{Lambert4} \\[12pt]
\int \frac{W\left(x\right)}{x} \ dx = \sum_{n = 1}^\infty \left(-1\right)^{n + 1} \frac{n^{n - 2}}{n!}\cdot x^n = W\left(x\right) + \frac{W^2\left(x\right)}{2} \label{Lambert5} \\[12pt]
\int \frac{1}{x} \int \frac{W\left(x\right)}{x} \ dx^2 = \sum_{n = 1}^\infty \left(-1\right)^{n + 1} \frac{n^{n - 3}}{n!}\cdot x^n = W\left(x\right) + \frac{3W^2\left(x\right)}{4} + \frac{W^3\left(x\right)}{6}. \qquad \label{Lambert6}
\end{IEEEeqnarray} \\
\section[Elliptic Integrals and Jacobian Elliptic Functions]{Elliptic Integrals and Jacobian Elliptic Functions \label{EllipticFunctionsAppendix}}
This appendix is a brief reminder of the definitions and some basic properties of elliptic integrals and Jacobian elliptic functions that we have used in our paper. Our conventions are similar to those of Abramowitz-Stegun \cite{AbramowitzStegun65}. \\[12pt]
\underline{Jacobian Elliptic Functions.}
\begin{IEEEeqnarray}{ll}
u \equiv \int_0^\varphi \frac{d\theta}{\left(1 - m \sin^2\theta\right)^{1/2}} \,, \quad & \varphi \equiv am(u | m) \,, \quad \Delta(\varphi) \equiv (1 - \sin^2\theta)^{1/2} \equiv dn(u | m) \nonumber \\
& x = \sin\varphi \equiv sn(u | m) \,, \quad \cos\varphi \equiv cn(u | m). \nonumber
\end{IEEEeqnarray}
\underline{Elliptic Integral of the First Kind.}
\begin{IEEEeqnarray}{l}
\mathbb{F}\left(\varphi \big| m\right) \equiv \int_0^\varphi \left(1 - m\,\sin^2\theta\right)^{-1/2} \, d\theta = \int_0^x \left[\left(1 - t^2\right)\left(1 - m\,t^2\right)\right]^{-1/2} \, dt = u \qquad \label{EllipticF1} \\[12pt]
\mathbb{K}\left(m\right) \equiv \mathbb{F}\left(\frac{\pi}{2} \Big| m\right) = \frac{\pi}{2} \cdot {_2\mathcal{F}_1}\left[\frac{1}{2},\frac{1}{2};1;m\right] \quad \text{(complete)} \label{EllipticK1}
\end{IEEEeqnarray}
\begin{IEEEeqnarray}{ll}
\mathbb{K}\left(m\right) &= \frac{\pi}{2} \cdot \sum_{n = 0}^\infty \left(\frac{(2n - 1)!!}{(2n)!!}\right)^2 m^n = \nonumber \\[12pt]
& = \frac{\pi}{2}\cdot\left[1 + \left(\frac{1}{2}\right)^2 m + \left(\frac{1 \cdot 3}{2 \cdot 4}\right)^2 m^2 + \left(\frac{1 \cdot 3 \cdot 5}{2 \cdot 4 \cdot 6}\right)^2 m^3 + \ldots \right] \,, \quad |m| < 1 \qquad
\end{IEEEeqnarray}
\begin{IEEEeqnarray}{ll}
\mathbb{K}\left(m\right) & = \frac{1}{2 \pi} \cdot \sum_{n = 0}^\infty \left(\frac{\Gamma(n + 1/2)}{n!}\right)^2 \left[2\psi(n + 1) - 2\psi(n + \frac{1}{2}) - \ln(1 - m)\right] \, (1 - m)^n = \nonumber \\[12pt]
& = \sum_{n = 0}^\infty \left(\frac{\left(2n - 1\right)!!}{\left(2n\right)!!}\right)^2 \left[\psi(n + 1) - \psi(n + \frac{1}{2}) - \frac{1}{2}\ln(1 - m)\right] \, (1 - m)^n \,, \nonumber \\[12pt]
& |1 - m| < 1, \qquad
\end{IEEEeqnarray}
where $\psi(z) \equiv \Gamma'(z)/\Gamma(z)$ is the psi (digamma) function. \\[18pt]
\underline{Elliptic Integral of the Second Kind.} \\
\begin{IEEEeqnarray}{l}
\mathbb{E}\left(\varphi \big| m\right) \equiv \int_0^\varphi \left(1 - m\,\sin^2\theta\right)^{1/2} \, d\theta = \int_0^x \left(1 - t^2\right)^{-1/2} \left(1 - m\,t^2\right)^{1/2} \, dt \qquad \label{EllipticE1} \\[12pt]
\mathbb{E}\left(m\right) \equiv \mathbb{E}\left(\frac{\pi}{2} \Big| m\right) = \frac{\pi}{2} \cdot {_2\mathcal{F}_1}\left[-\frac{1}{2},\frac{1}{2};1;m\right] \quad \text{(complete)} \label{EllipticE2}
\end{IEEEeqnarray}
\begin{IEEEeqnarray}{ll}
\mathbb{E}\left(m\right) & = - \frac{\pi}{2} \cdot \sum_{n = 0}^\infty \left(\frac{(2n - 1)!!}{(2n)!!}\right)^2 \frac{m^n}{2n - 1} = \nonumber \\[12pt]
& = \frac{\pi}{2} \cdot \left[1 - \left(\frac{1}{2}\right)^2 \frac{m}{1} - \left(\frac{1 \cdot 3}{2 \cdot 4}\right)^2 \frac{m^2}{3} - \left(\frac{1 \cdot 3 \cdot 5}{2 \cdot 4 \cdot 6}\right)^2 \frac{m^3}{5} + \ldots \right] \,, \quad |m| < 1 \qquad
\end{IEEEeqnarray}
\begin{IEEEeqnarray}{lll}
\mathbb{E}\left(m\right) & = 1 &- \frac{1}{2 \pi} \cdot \sum_{n = 0}^\infty \frac{\Gamma(n + 1/2) \Gamma(n + 3/2)}{n!(n + 1)!} \bigg[\ln(1 - m) + \psi(n + 1/2) + \psi(n + 3/2) - \nonumber \\[12pt]
&& - \psi(n + 1) - \psi(n + 2)\bigg] \, (1 - m)^{n + 1} = \nonumber \\[12pt]
& = 1 &+ \frac{1}{2 \pi} \cdot \sum_{n = 0}^\infty \frac{\Gamma(n + 1/2) \Gamma(n + 3/2)}{n!(n + 1)!} \bigg[2\psi(n + 1) - 2\psi(n + 1/2) - \ln(1 - m) - \nonumber \\[12pt]
&& - \frac{1}{\left(n + 1\right)\left(2n + 1\right)}\bigg] \, (1 - m)^{n + 1} \,, \quad |1 - m| < 1. \qquad
\end{IEEEeqnarray} \\
\section[Short and Slow Strings]{Short and Slow Strings \label{Short-Slow-StringsAppendix}}
In this appendix we summarize our computations on short and slow GKP strings. We have the following cases. \\[12pt]
$\bullet$ $\mathbb{R}\times\text{S}^2$ folded strings ($\omega > 1$): for short folded strings on S$^2$ ($\omega \rightarrow \infty$) the expansions of energy \eqref{GKPIIEnergy2} and spin \eqref{GKPIISpin2} in terms of the angular frequency $\omega$ assume the following forms (cf. appendix \ref{EllipticFunctionsAppendix}):
\begin{IEEEeqnarray}{l}
E = \sqrt{\lambda} \cdot \sum_{n = 0}^{\infty} \left(\frac{\left(2n - 1\right)!!}{\left(2n\right)!!}\right)^2 \frac{1}{\omega^{2n + 1}} \qquad \label{ShortStringIIEnergy1} \\[6pt]
J = \sqrt{\lambda} \cdot \sum_{n = 1}^{\infty} \left(\frac{\left(2n - 1\right)!!}{\left(2n\right)!!}\right)^2 \frac{2n}{2n - 1} \frac{1}{\omega^{2n}}. \qquad \label{ShortStringIISpin}
\end{IEEEeqnarray}
Series \eqref{ShortStringIISpin} can be inverted with a symbolic computations program, then the inverse spin function $x = x\left(\mathcal{J}\right)$ may be inserted into the expression for energy \eqref{ShortStringIIEnergy1}, leading to $\mathcal{E} = \mathcal{E}\left(\mathcal{J}\right)$. The results are: \\
\small\begin{IEEEeqnarray}{ll}
x = 1 - \frac{4\mathcal{J}}{\pi} + \frac{6\mathcal{J}^2}{\pi^2} - \frac{3\mathcal{J}^3}{\pi^3} - \frac{5\mathcal{J}^4}{4\pi^4} + \frac{9\mathcal{J}^5}{16\pi^5} + \frac{21\mathcal{J}^6}{16\pi^6} + \frac{35\mathcal{J}^7}{64\pi^7} - \frac{459\mathcal{J}^8}{512\pi^8} - \frac{5.835\mathcal{J}^9}{4.096\pi^9} - \ldots \label{ShortStringII-InverseSpin} \\[12pt]
\mathcal{E} = \pi^{1/2}\mathcal{J}^{1/2} + \frac{\mathcal{J}^{3/2}}{4\,\pi^{1/2}} + \frac{3\,\mathcal{J}^{5/2}}{32\,\pi^{3/2}} + \frac{\mathcal{J}^{7/2}}{128\,\pi^{5/2}} - \frac{61\,\mathcal{J}^{9/2}}{2.048\,\pi^{7/2}} - \frac{201\,\mathcal{J}^{11/2}}{8.192\,\pi^{9/2}} + \frac{199\,\mathcal{J}^{13/2}}{65.536\,\pi^{11/2}} + \ldots \qquad \label{ShortStringIIEnergy2}
\end{IEEEeqnarray} \normalsize \\
The latter may also be written as follows: \\
\small\begin{IEEEeqnarray}{ll}
E &= \left(2 \, \sqrt{\lambda} \, J\right)^{1/2} \cdot \left[1 + \frac{J}{8 \sqrt{\lambda}} + \frac{3 J^2}{128\,\lambda} + \frac{J^3}{1\,024\,\lambda^{3/2}} - \frac{61 J^4}{32\,768\,\lambda^2} - \frac{201 J^5}{262\,144\,\lambda^{5/2}} + O\left(\frac{J^6}{\lambda^3}\right)\right]. \qquad \label{ShortStringIIEnergy3}
\end{IEEEeqnarray} \normalsize \\
$\bullet$ $\mathbb{R}\times\text{S}^2$ circular strings ($\omega < 1$): slow circular strings on S$^2$ ($\omega \rightarrow 0^+$) have the following series for the energy \eqref{GKPIICircularEnergy1} and spin \eqref{GKPIICircularSpin1}, in terms of the angular frequency $\omega$:
\begin{IEEEeqnarray}{l}
E = \sqrt{\lambda} \cdot \sum_{n = 0}^{\infty} \left(\frac{\left(2n - 1\right)!!}{\left(2n\right)!!}\right)^2 \omega^{2n} \qquad \label{ShortStringIICircularEnergy1} \\[6pt]
J = \sqrt{\lambda} \cdot \sum_{n = 1}^{\infty} \left(\frac{\left(2n - 1\right)!!}{\left(2n\right)!!}\right)^2 \frac{2n}{2n - 1} \cdot \omega^{2n-1}. \qquad \label{ShortStringIICircularSpin}
\end{IEEEeqnarray}
Again, series \eqref{ShortStringIICircularSpin} is reverted for $\omega = \omega\left(J\right)$ and then the inverse spin function $\widetilde{x} \equiv 1 - \omega^2 = \widetilde{x}\left(\mathcal{J}\right)$ is plugged into the expression for energy \eqref{ShortStringIICircularEnergy1}, leading to: \\
\small\begin{IEEEeqnarray}{ll}
\widetilde{x} = 1 - \frac{16\,\mathcal{J}^2}{\pi^2} + \frac{192\,\mathcal{J}^4}{\pi^4} - \frac{2112\,\mathcal{J}^6}{\pi^6} + \frac{22.400\,\mathcal{J}^8}{\pi^8} - \frac{233.088\,\mathcal{J}^{10}}{\pi^{10}} + \frac{2.397.696\,\mathcal{J}^{12}}{\pi^{12}} - \ldots \label{ShortStringII-CircularInverseSpin} \\[12pt]
\mathcal{E} = \frac{\pi}{2} + \frac{2\,\mathcal{J}^2}{\pi} - \frac{6\,\mathcal{J}^4}{\pi^3} + \frac{32\,\mathcal{J}^6}{\pi^5} - \frac{206\,\mathcal{J}^8}{\pi^7} + \frac{1464\,\mathcal{J}^{10}}{\pi^9}-\frac{11.064\,\mathcal{J}^{12}}{\pi^{11}} + \frac{87.200\,\mathcal{J}^{14}}{\pi^{13}} - \ldots \qquad \label{ShortStringIICircularEnergy2}
\end{IEEEeqnarray} \normalsize \\
\eqref{ShortStringIICircularEnergy2} can also be written as follows: \\
\small\begin{IEEEeqnarray}{ll}
E &= \sqrt{\lambda} \cdot \left[1 + \frac{J^2}{\lambda} - \frac{3\, J^4}{4\,\lambda^2} + \frac{J^6}{\lambda^3} - \frac{103\,J^8}{64\,\lambda^4} + \frac{183\,J^{10}}{64\,\lambda^5} - \frac{1383\,J^{12}}{256\,\lambda^6} + \frac{2725\,J^{14}}{256\,\lambda^7} - O\left(\frac{J^{16}}{\lambda^8}\right)\right]. \qquad \label{ShortStringIICircularEnergy3}
\end{IEEEeqnarray} \normalsize \\
$\bullet$ $\text{AdS}_3$ folded strings ($\omega > 1$): for the short-string limit of AdS$_3$ strings, we may expand expressions \eqref{GKPIEnergy2}-\eqref{GKPISpin2} around $\omega \rightarrow \infty$ and obtain the corresponding short-string series (cf. appendix \ref{EllipticFunctionsAppendix}):
\begin{IEEEeqnarray}{l}
E = \sqrt{\lambda} \cdot \sum_{n = 0}^{\infty} \left(\frac{\left(2n - 1\right)!!}{\left(2n\right)!!}\right)^2 \frac{2n + 1}{\omega^{2n + 1}} \qquad \label{ShortStringIEnergy1} \\[6pt]
S = \sqrt{\lambda} \cdot \sum_{n = 1}^{\infty} \left(\frac{\left(2n - 1\right)!!}{\left(2n\right)!!}\right)^2 \frac{2n}{\omega^{2n}}, \qquad \label{ShortStringISpin}
\end{IEEEeqnarray}
where, in order to obtain series \eqref{ShortStringIEnergy1}, the identity
\begin{IEEEeqnarray}{c}
\left(2n + 1\right) \, \left(\frac{\left(2n - 1\right)!!}{\left(2n\right)!!}\right)^2 + \sum_{k = 0}^{n} \frac{1}{2k - 1} \left(\frac{\left(2k - 1\right)!!}{\left(2k\right)!!}\right)^2 = 0
\end{IEEEeqnarray}
was used. Observe that the short S$^2$-string coefficients of the energy series \eqref{ShortStringIIEnergy1} differ from the corresponding AdS$_3$ ones \eqref{ShortStringIEnergy1} by a factor of $2n + 1$, $n = 0,\, 1,\, \ldots$, while the coefficients of the angular momentum $J$ in \eqref{ShortStringIISpin} differ from those of the spin $S$ in \eqref{ShortStringISpin} by $1/2n - 1$. This is related to the fact the the corresponding functions \eqref{GKPIEnergy2}, \eqref{GKPISpin2} can appropriately be taken by differentiation/integration of \eqref{GKPIIEnergy2}, \eqref{GKPIISpin2}. Using series reversion in Mathematica, we may also obtain expressions for the inverse spin function $x = x\left(\mathcal{S}\right)$ and energy $\mathcal{E} = \mathcal{E}\left(\mathcal{S}\right)$ as functions of the spin $\mathcal{S}$: \\
\small\begin{IEEEeqnarray}{ll}
x = 1 - \frac{4\,\mathcal{S}}{\pi} + \frac{18\,\mathcal{S}^2}{\pi^2} - \frac{87\,\mathcal{S}^3}{\pi^3} + \frac{1.765\,\mathcal{S}^4}{4\,\pi^4} - \frac{37.071\,\mathcal{S}^5}{16\,\pi^5} + \frac{199.815\,\mathcal{S}^6}{16\,\pi^6} - \frac{4.397.017\,\mathcal{S}^7}{64\,\pi^7} + \ldots \qquad \label{ShortStringI-InverseSpin} \\[12pt]
\mathcal{E} = \pi^{1/2}\mathcal{S}^{1/2} + \frac{3\,\mathcal{S}^{3/2}}{4\,\pi^{1/2}} - \frac{21\,\mathcal{S}^{5/2}}{32\,\pi^{3/2}} + \frac{187\,\mathcal{S}^{7/2}}{128\,\pi^{5/2}} - \frac{9.261\,\mathcal{S}^{9/2}}{2.048\,\pi^{7/2}} + \frac{136.245\,\mathcal{S}^{11/2}}{8.192\,\pi^{9/2}} - \ldots \qquad \label{ShortStringIEnergy2}
\end{IEEEeqnarray} \normalsize \\
The 't Hooft coupling dependence of the last expression may be restored as follows: \\
\small\begin{IEEEeqnarray}{ll}
E & = \left(2 \, \sqrt{\lambda} \, S\right)^{1/2} \left[1 + \frac{3\,S}{8\,\sqrt{\lambda}} - \frac{21\,S^2}{128\,\lambda} + \frac{187\,S^3}{1\,024\,\lambda^{3/2}} - \frac{9\,261\,S^4}{32\,768\,\lambda^2} + \frac{136\,245\,S^5}{262\,144\,\lambda^{5/2}} - O\left(\frac{S^6}{\lambda^3}\right)\right]. \qquad \label{ShortStringIEnergy3}
\end{IEEEeqnarray} \normalsize \\
\section[Long and Fast Strings]{Long and Fast Strings \label{Long-Fast-StringsAppendix}}
This appendix contains some of our symbolic computations on long and fast GKP strings. Just as for short/slow strings in the preceding appendix, the inverse spin functions $x\left(\mathcal{J}\right)$ and $x\left(\mathcal{S}\right)$ of long/fast strings on $\mathbb{R}\times\text{S}^2$ and $\text{AdS}_3$ have been obtained by inverting the corresponding series for spins \eqref{GKPIISpin3}, \eqref{GKPIICircularSpin2} and \eqref{GKPISpin4} with Mathematica. Then the results are plugged into the series of the anomalous dimensions $\mathcal{E} = \mathcal{E}\left(x\right)$, leading to the expressions for $\mathcal{E}\left(\mathcal{J}\right)$ and $\mathcal{E}\left(\mathcal{S}\right)$. Only the first few terms of each series are presented here. All of these results agree with the analytic formulas and series coefficients that were derived in our paper and we summarized in section \ref{StrongSummary}. We have addressed the following cases. \\[12pt]
$\bullet$ Folded string on $\mathbb{R}\times\text{S}^2$ ($\omega > 1$).\footnote{By making the transformation
\begin{IEEEeqnarray}{l}
\mathcal{S} \equiv \frac{1}{16} \, e^{2\mathcal{J} + 2} \Leftrightarrow \mathcal{J} = \frac{1}{2} \left(\ln \mathcal{S} + 4\ln2 - 2\right) \label{ReciprocityTransform}
\end{IEEEeqnarray}
the inverse spin function and the anomalous dimensions of strings spinning in $\mathbb{R}\times\text{S}^2$, assume a form that resembles the corresponding formula for strings in AdS$_3$ and permits comparisons between the two. Interestingly, and in contradistinction with the AdS case, all $\ln^n \mathcal{S} / \mathcal{S}^n$ terms are absent.} \\
\footnotesize\begin{IEEEeqnarray}{ll}
x = & \; 16\,e^{-2\mathcal{J} - 2} - 128 \left(\mathcal{J} + 1\right)\,e^{-4\mathcal{J} - 4} + 64 \left(24\mathcal{J}^2 + 34\mathcal{J} + 15\right)\,e^{-6\mathcal{J} - 6} - \frac{512}{3}  (128\mathcal{J}^3 + 216\mathcal{J}^2 + 153\mathcal{J} + \nonumber \\[12pt]
&+ 42)\,e^{-8\mathcal{J} - 8} + \frac{32}{3} (32.000\mathcal{J}^4 + 60.800\mathcal{J}^3 + 54.960\mathcal{J}^2 + 25.452\mathcal{J} + 4989)\,e^{-10\mathcal{J} - 10} - \frac{512}{5}(55.296 \mathcal{J}^5 + \nonumber \\[12pt]
& + 115.200 \mathcal{J}^4 + 122.400 \mathcal{J}^3 + 74.460 \mathcal{J}^2 + 25.480 \mathcal{J} + 3855) \,e^{-12\mathcal{J} - 12} + \ldots \label{Exact_Value_of_xII1}
\end{IEEEeqnarray} \normalsize

\footnotesize\begin{IEEEeqnarray}{lll}
\mathcal{E} - \mathcal{J} = 1 &- 4 e^{-2\mathcal{J} - 2} + 4 \left(4\mathcal{J} - 1\right) \, e^{-4\mathcal{J} - 4} - 32 \left(4\mathcal{J}^2 - \mathcal{J} + 1\right) \, e^{-6\mathcal{J} - 6} + \frac{4}{3}(1024\mathcal{J}^3 - 192\mathcal{J}^2 + 456\mathcal{J} - \nonumber \\[12pt]
& - 63) \, e^{-8\mathcal{J} - 8} - \frac{8}{3}(6400\mathcal{J}^4 - 640\mathcal{J}^3 + 3888\mathcal{J}^2 - 660\mathcal{J} + 279) \, e^{-10\mathcal{J} - 10} + \frac{32}{5} (36.864 \mathcal{J}^5 + \nonumber \\[12pt]
& + 27.840 \mathcal{J}^3 - 4040 \mathcal{J}^2 + 4160 \mathcal{J} - 405) \, e^{-12\mathcal{J} - 12} - \ldots \qquad \label{MathematicaAnomalousDimensionsII1}
\end{IEEEeqnarray} \normalsize \\[6pt]
$\bullet$ Circular string on $\mathbb{R}\times\text{S}^2$ ($\omega < 1$). \\
\footnotesize\begin{IEEEeqnarray}{ll}
\widetilde{x} = & \; 16\,e^{-2\mathcal{J} - 2} + 128 \left(\mathcal{J} - 1\right)\,e^{-4\mathcal{J} - 4} + 192 \left(8\mathcal{J}^2 - 10\mathcal{J} + 5\right)\,e^{-6\mathcal{J} - 6} + \frac{512}{3}  (128\mathcal{J}^3 - 168\mathcal{J}^2 + 129\mathcal{J} - \nonumber \\[12pt]
& - 42)\,e^{-8\mathcal{J} - 8} + \frac{32}{3} (32.000\mathcal{J}^4 - 41.600\mathcal{J}^3 + 39.600\mathcal{J}^2 - 20.628\mathcal{J} + 4989)\,e^{-10\mathcal{J} - 10} + \frac{1536}{5}(18.432 \mathcal{J}^5 \nonumber \\[12pt]
& - 23.040 \mathcal{J}^4 + 25.440 \mathcal{J}^3 - 16.460 \mathcal{J}^2 + 6.720 \mathcal{J} - 1285) \,e^{-12\mathcal{J} - 12} + \ldots \label{Exact_Value_of_xII2}
\end{IEEEeqnarray} \normalsize \\
\footnotesize\begin{IEEEeqnarray}{ll}
\mathcal{E} - \mathcal{J} = & 1 + 4 e^{-2\mathcal{J} - 2} + 4 \left(4\mathcal{J} - 1\right) \, e^{-4\mathcal{J} - 4} + 32 \left(4\mathcal{J}^2 - \mathcal{J} + 1\right) \, e^{-6\mathcal{J} - 6} + \frac{4}{3}(1024\mathcal{J}^3 - 192\mathcal{J}^2 + 456\mathcal{J} - \nonumber \\[12pt]
& - 63) \, e^{-8\mathcal{J} - 8} + \frac{8}{3}(6400\mathcal{J}^4 - 640\mathcal{J}^3 + 3888\mathcal{J}^2 - 660\mathcal{J} + 279) \, e^{-10\mathcal{J} - 10} + \frac{32}{5} (36.864 \mathcal{J}^5 + \nonumber \\[12pt]
& + 27.840 \mathcal{J}^3 - 4040 \mathcal{J}^2 + 4160 \mathcal{J} - 405) \, e^{-12\mathcal{J} - 12} + \ldots \label{MathematicaAnomalousDimensionsII2}
\end{IEEEeqnarray} \normalsize \\[6pt]
$\bullet$ Folded string on $\text{AdS}_3$ ($\omega > 1$). \\
\footnotesize\begin{IEEEeqnarray}{ll}
x = \frac{1}{\mathcal{S}} &- \Big[\frac{1}{4}\ln\mathcal{S} + \Big(\ln2 + \frac{1}{4}\Big)\Big]\frac{1}{\mathcal{S}^2} + \Big[\frac{1}{16}\ln^2\mathcal{S} + \Big(\frac{\ln2}{2} + \frac{1}{32}\Big) \ln\mathcal{S} + \Big(\ln^2 2 + \frac{\ln2}{8} + \frac{3}{64}\Big)\Big]\frac{1}{\mathcal{S}^3} - \Big[\frac{1}{64}\ln^3\mathcal{S} + \nonumber \\[12pt]
& + \Big(\frac{3\ln2}{16} - \frac{1}{64}\Big)\ln^2\mathcal{S} + \Big(\frac{3\ln^2 2}{4} - \frac{\ln2}{8} + \frac{3}{128}\Big)\ln\mathcal{S} + \Big(\ln^3 2 - \frac{\ln^2 2}{4} + \frac{3\ln2}{32}\Big)\Big]\frac{1}{\mathcal{S}^4} + \ldots \label{Exact_Value_of_xI1}
\end{IEEEeqnarray} \normalsize \\
\footnotesize\begin{IEEEeqnarray}{ll}
\gamma = &\frac{1}{2}\ln\mathcal{S} + \Big[2\ln2 - \frac{1}{2}\Big] + \Big[\frac{1}{8}\ln\mathcal{S} + \left(\frac{\ln2}{2} - \frac{1}{8}\right)\Big]\frac{1}{\mathcal{S}} - \Big[\frac{1}{64}\ln^2\mathcal{S} + \left(\frac{\ln2}{8} - \frac{9}{128}\right)\ln\mathcal{S} + \bigg(\frac{\ln^2 2}{4} - \frac{9\ln2}{32} + \nonumber \\[12pt]
& + \frac{5}{128}\bigg)\Big]\frac{1}{\mathcal{S}^2} + \Big[\frac{1}{384}\ln^3\mathcal{S} + \left(\frac{\ln2}{32} - \frac{3}{128}\right)\ln^2\mathcal{S} + \left(\frac{\ln^2 2}{8} - \frac{3\ln2}{16} + \frac{11}{256}\right)\ln\mathcal{S} + \bigg(\frac{\ln^3 2}{6} - \frac{3\ln^2 2}{8} + \nonumber \\[12pt]
& + \frac{11 \ln2}{64} - \frac{7}{384}\bigg)\Big]\frac{1}{\mathcal{S}^3} - \Big[ \frac{1}{2048}\ln^4\mathcal{S} + \left(\frac{\ln2}{128} - \frac{43}{6144}\right)\ln^3\mathcal{S} + \left(\frac{3\ln^2 2}{64} - \frac{43 \ln2}{512} + \frac{51}{2048}\right)\ln^2\mathcal{S} + \nonumber \\[12pt]
& + \Big(\frac{\ln^3 2}{8} - \frac{43 \ln^2 2}{128} + \frac{51 \ln2}{256} - \frac{937}{32.768}\Big)\ln\mathcal{S} + \left(\frac{\ln^4 2}{8} - \frac{43 \ln^3 2}{96} + \frac{51 \ln^2 2}{128} - \frac{937 \ln2}{8192} + \frac{1919}{196.608}\right)\Big]\cdot \nonumber \\[12pt]
& \cdot\frac{1}{\mathcal{S}^4} + \ldots \label{MathematicaAnomalousDimensionsI1}
\end{IEEEeqnarray} \normalsize
\bibliographystyle{JHEP}
\bibliography{Bibliography}
\end{document}